\documentclass[aps,prd,preprint,floatfix]{revtex4}
\usepackage{epsfig}
\usepackage{dcolumn}

\begin{document}


\preprint{\rightline{ANL-HEP-PR-03-022}}

\title{The pseudo-Goldstone spectrum of 2-colour QCD at finite density.}

\author{J.~B.~Kogut and D.~Toublan}
\address{Dept. of Physics, University of Illinois, 1110 West Green Street,
Urbana, IL 61801-3080, USA}
\author{D.~K.~Sinclair}
\address{HEP Division, Argonne National Laboratory, 9700 South Cass Avenue,
Argonne, IL 60439, USA}

\begin{abstract}
We examine the spectrum of 2-colour lattice QCD with 4 continuum flavours at a
finite chemical potential ($\mu$) for quark-number, on a $12^3 \times 24$
lattice. First we present evidence that the system undergoes a transition to a
state with a diquark condensate, which spontaneously breaks quark number at
$\mu=m_\pi/2$, and that this transition is mean field in nature. We then
examine the 3 states that would be Goldstone bosons at $\mu=0$ for zero Dirac
and Majorana quark masses. The predictions of chiral effective Lagrangians
give a good description of the behaviour of these masses for $\mu < m_\pi/2$.
Except for the heaviest of these states, these predictions diverge from our
measurements, once $\mu$ is significantly greater than $m_\pi/2$. However, the
qualitative behaviour of these masses, indicates that the physics is very
similar to that predicted by these effective Lagrangians, and there is some
indication that at least part of these discrepancies is due to saturation, a
lattice artifact.
\end{abstract}

\maketitle

\section{Introduction}

Recently there has been a renewed interest in the properties of nuclear matter
--- hadronic matter at finite non-zero baryon number (and isospin) density
\cite{Shuryak,Wilczek}. Much of this interest comes from a reevaluation of the
old idea that quark pairs might condense giving rise to a transition to a
colour superconducting state at high baryon-number density
\cite{Love,Barrois}. These newer studies indicate that the energies associated
with this transition are much larger than the original estimates so that they
could have a significant effect on the equation-of-state of nuclear matter.
Unfortunately, adding a finite chemical potential for quark/baryon number to
the Euclidean QCD action renders the fermion determinant complex which
precludes the naive application of standard lattice simulation methods. (Some
advances have been made allowing studies at small chemical potential and high
temperatures, but their ranges of applicability are limited \cite{mu-T}.) For
this reason people have turned to the study of models which have some of the
properties of QCD at finite baryon number but have real, positive fermion
determinants, allowing lattice simulations.

One such model is 2-colour QCD with fundamental quarks. With $SU(2)_{colour}$,
the quarks and anti-quarks are in the same representation, leading to a fermion
determinant which remains real and positive in the presence of a quark-number
chemical potential, $\mu$. It is expected that, at $\mu=m_\pi/2$, this model
exhibits a phase transition to a state with a colourless diquark condensate
which breaks quark number, and associated Goldstone bosons. Thus this
condensed phase is a superfluid rather than a superconductor. Chiral effective
Lagrangians have been used to predict the phase structure of this theory
\cite{Toublan,KSTVZ,STV1,STV2,SSS} as have random-matrix models 
\cite{Vanderheyden}. 

This predicted phase structure including the mean-field nature of the
transition has been observed in lattice simulations 
\cite{HKLM,HKMS,KTS1,KTS2,KTS3,aadgg,t+mu}. However, the only
simulations in which it was possible to study the masses of potential Goldstone
bosons were performed at a quark mass so large that it was difficult
to measure those excitations which were significantly more massive than the
pion \cite{HKMS}. It was also unclear at such large quark masses if the masses
of these pseudo-Goldstone modes were well separated from the rest of the hadron
spectrum. Since these earlier simulations were performed at a quark mass of
$m=0.1$, we have performed new simulations at the same $\beta$(1.5) and quark
mass $m=0.025$ which should halve the pion mass $m_\pi$. These simulations
were performed on a $12^3 \times 24$ lattice again with the quark-number
symmetry breaking parameter $\lambda=0.1 m$ and $\lambda=0.2 m$ to keep the
explicit symmetry breaking, which depends on $\lambda/m$, small. We have also
performed simulations on a smaller ($8^4$) lattice to give us some indication
of the magnitude of finite size effects.

In these simulations which had a single staggered quark field corresponding to
4 continuum quark flavours, the lattice flavour symmetry at $m=\lambda=\mu=0$
is $U(2)$. When this breaks spontaneously, $U(2) \rightarrow U(1)$ giving rise
to 3 Goldstone bosons. We study the behaviour of these 3 states for $m \ne 0$,
$\mu \ne 0$ and $\lambda << m$, and compare with the leading-order predictions
from chiral effective Lagrangians. For $\mu < m_\pi/2$, the agreement with
these predictions is excellent. As $\mu$ is increased beyond $m_\pi/2$, the
mass of the excitation which would be a Goldstone boson for $\lambda=0$ and
that of the pion, lie consistently below the predictions from tree-level
effective Lagrangians. Part of this difference might be due to higher order
corrections in the effective Lagrangian calculations, which cannot be
characterized by a single parameter in addition to those of tree-level
chiral perturbation theory. However, at least some of this `discrepancy'
appears to be due to the fact that, at high densities, the fermionic
constituents of these excitations are revealed, indicating that we are beyond
the reach of chiral perturbation theory. However, the high density behaviour
of the lattice theory is strongly affected by saturation, a lattice artifact,
so it is unclear how much of this would survive to the continuum.

In section~2 we review the expected pattern of symmetry breaking for lattice
2-colour QCD with 1 fundamental staggered quark field (4 flavours), and the
(pseudo)-Goldstone spectrum associated with this breaking. Throughout our
analysis we compare the lattice results to analytical calculations derived
from effective models. We use a non-linear-sigma model (chiral perturbation
theory) at leading order similar to that described in \cite{KSTVZ}, and a
linear sigma model that models some of the higher order corrections to
chiral perturbation theory. We present the linear sigma model effective
Lagrangian we use to fit our diquark condensates, and its predictions for the
pseudo-Goldstone spectrum in section~3, while, for completeness, the
non-linear sigma model is presented in Appendix~A. Section~4 describes our
simulations, presents our results for the condensates and the evidence for
mean-field scaling. In section~5 we present our measurements of the spectrum
of this theory, and its comparison with the predictions of effective
Lagrangian analyses. Our conclusions are presented in section~6.

\section{Symmetry breaking in 2-colour lattice QCD}

The staggered fermion action for 2-colour lattice QCD with 1-staggered fermion,
i.e. 4 continuum flavours in the fundamental representation of the colour
group is:
\begin{equation}
S_f = \sum_{sites}\left\{\bar{\chi}[D\!\!\!\!/\,(\mu) + m]\chi
+ \frac{1}{2}\lambda[\chi^T\tau_2\chi + \bar{\chi}\tau_2\bar{\chi}^T]\right\}
\label{eqn:lagrangian}
\end{equation}
where $D\!\!\!\!/\,(\mu)$ is the normal staggered covariant finite difference
operator with links in the $+t$ direction multiplied by $e^\mu$ and those in
the $-t$ direction by $e^{-\mu}$. What follows summarizes the analysis of
the symmetries of this theory presented in detail in \cite{HKLM}

At $\mu=m=\lambda=0$, this action has a $U(2)$ flavour symmetry which breaks
spontaneously to $U(1)$. If it breaks forming a chiral condensate, 
$\langle\bar{\chi}\chi\rangle$, there will be 3 broken $U(2)$ generators and
3 Goldstone bosons, namely;
\begin{eqnarray}
\label{eqn:chisb}
{\bf 1}  &\Longrightarrow& \bar{\chi}\epsilon\chi     \nonumber            \\
\sigma_1 &\Longrightarrow& \chi^T\tau_2\chi - \bar{\chi}\tau_2\bar{\chi}^T \\
\sigma_2 &\Longrightarrow& \chi^T\tau_2\chi + \bar{\chi}\tau_2\bar{\chi}^T
\nonumber.
\end{eqnarray}
If it breaks forming a diquark condensate,
$\frac{1}{2}\langle\chi^T\tau_2\chi + \bar{\chi}\tau_2\bar{\chi}^T\rangle$, 
there will again be 3 broken generators and 3 Goldstone bosons,
\begin{eqnarray}
\label{eqn:qsb}
{\bf 1}  &\Longrightarrow& \chi^T\tau_2\epsilon\chi +
                           \bar{\chi}\tau_2\epsilon\bar{\chi}^T \nonumber \\
\sigma_2 &\Longrightarrow& \bar{\chi}\chi                                 \\
\sigma_3 &\Longrightarrow& \chi^T\tau_2\chi - \bar{\chi}\tau_2\bar{\chi}^T
\nonumber.
\end{eqnarray} 
If $\mu \ne 0$ and $m=0$, the staggered fermion action
\ref{eqn:lagrangian} is invariant under 
$U(1)\times U(1)$, which is completely spontaneously broken by the
diquark condensate. Therefore, only the 2 diquark states in
equation~\ref{eqn:qsb} remain Goldstone bosons.

When $m \ne 0$, the chiral condensate forms and the 3 Goldstone bosons of
equation~\ref{eqn:chisb} gain equal masses given by PCAC. As $\mu$ is
increased from zero, the mass of the pion created by $\bar{\chi}\epsilon\chi$
remains constant, since it has zero quark number and does not feel the effect
of the chemical potential. The energy and hence the mass of the
diquark state
created by $\bar{\chi}\tau_2\bar{\chi}^T$ is increased to $m_\pi+2\mu$ for
diquarks propagating in the $+t$ direction. The mass of the forward
propagating anti-diquark is decreased to $m_\pi-2\mu$. 



This latter mass
vanishes at $\mu=m_\pi/2$, and it becomes a true Goldstone boson. This heralds
the phase transition to a state in which quark-number is spontaneously broken
by a diquark condensate,
$\frac{1}{2}\langle\chi^T\tau_2\chi + \bar{\chi}\tau_2\bar{\chi}^T\rangle$.
The Goldstone boson is created by the orthogonal linear combination,
$\frac{1}{2}(\chi^T\tau_2\chi - \bar{\chi}\tau_2\bar{\chi}^T)$,
which we note is just that Goldstone mode which is common to 
equations~\ref{eqn:chisb} and \ref{eqn:qsb}.

For $\mu > m_\pi/2$, it is useful to introduce the concept of a total 
condensate $\Sigma_c$ such that:
\begin{equation}
\label{eqn:rot1}
\langle\bar{\chi}\chi\rangle = \Sigma_c \cos \alpha,
\end{equation}
and
\begin{equation}
\label{eqn:rot2}
\frac{1}{2}\langle\chi^T\tau_2\chi + \bar{\chi}\tau_2\bar{\chi}^T\rangle
                             = \Sigma_c \sin \alpha.
\end{equation}
We then see that the heaviest of our 3 would-be Goldstone bosons will be that
created by the operator
\begin{equation}
\frac{1}{2}(\chi^T\tau_2\chi + \bar{\chi}\tau_2\bar{\chi}^T) \cos \alpha
                             - \bar{\chi}\chi \sin \alpha                ,
\label{eqn:scalar}
\end{equation}
which has zero vacuum expectation value. For $\mu$ just above the transition,
this is predominantly a diquark/anti-diquark state, while for $\mu$ large, it
is predominantly the scalar $\sigma/f_0$ meson, as expected since $m$ can be
neglected. This state will be heavy since the scalar meson is not a Goldstone
boson when $\mu > 0$, even when $m=0$. Finally, the pseudo-scalar
pseudo-Goldstone boson is created by the operator
\begin{equation}
\bar{\chi}\epsilon\chi \cos \bar{\alpha} - \frac{1}{2}(\chi^T\tau_2\epsilon\chi
             + \bar{\chi}\tau_2\epsilon\bar{\chi}^T) \sin \bar{\alpha}
\label{eqn:pi}
\end{equation}
where effective Lagrangians suggest $\bar{\alpha}=\alpha$, with
$\alpha$ introduced in (\ref{eqn:rot1}, \ref{eqn:rot2}). For $\mu$ small,
this state is predominantly a pion, while for large $\mu$ it is predominantly
a pseudoscalar diquark. Since this pseudoscalar diquark would be a Goldstone
boson if $m=0$, its mass should approach zero for large $\mu$ when $m$ can be
neglected.

When $\lambda \ne 0$, the remaining Goldstone boson becomes massive. For the
details of this $\lambda$ dependence and for the $\mu$ dependence of the
heavier, would-be Goldstone modes above the transition, we must turn to
effective Lagrangians and chiral perturbation theory.

\section{Linear Sigma Model Effective Lagrangian}

For small pion mass $m_\pi$, $\mu$ and $\lambda$, we should be able to use
chiral perturbation theory to parametrize the behaviour of the condensates,
quark-number density and pseudo-Goldstone spectrum of this theory, as was done
in \cite{Toublan,KSTVZ}. A reworking of this analysis for the symmetries of
the staggered quark action is presented in the appendix. 
However, we find that $m_\pi$, even for the lowest quark mass $m$ we
use, is too large for
tree-level chiral perturbation theory to give a quantitative description of
the physics of this system except a relatively low chemical potential. 
While next-to-leading order chiral perturbation
theory calculations have been performed \cite{STV1}, these do not yet include
spectrum calculations. Even the calculation of the order parameters has not
been extended beyond the neighbourhood of the critical point at
next-to-leading order.

We therefore introduce an alternative effective Lagrangian which incorporates
at tree level some of the properties expected from an all-orders chiral
perturbation theory calculation. First, it should have the same phase structure
and critical exponents (mean field) as tree-level chiral perturbation theory,
and a critical point at $\mu=m_\pi/2$. At $\lambda=0$, the spectrum of 
pseudo-Goldstone bosons for $\mu < m_\pi/2$ should be that predicted in the
previous section from fairly general arguments, and for $\mu > m_\pi/2$, it
should have one massless Goldstone boson. The magnitude of the total
condensate $\sqrt{\langle \bar{\chi} \chi \rangle^2 + (\frac12 \langle \chi^T
\tau_2 \chi + \bar{\chi} \tau_2 \bar{\chi}^T\rangle)^2}$ 
is independent of $\mu$ in tree-level chiral perturbation
theory, but depends on $\mu$ at next-to-leading order. Lattice results
indeed indicate that the total condensate increases when $\mu$
increases. Therefore the total condensate should be
allowed to vary. Such variation can be allowed if the
magnitude of the condensate becomes a dynamical field. With the chiral
perturbation theory Lagrangian, this would be a non-perturbative effect, since
it involves producing a bound state. Modifying our Lagrangian to explicitly
incorporate such excitations requires replacing the chiral Lagrangian which is
of the non-linear sigma model class by the corresponding linear sigma model
effective Lagrangian. Since we do not intend to use this Lagrangian beyond
tree level, we do not have to face the problems of trying to formulate a chiral
perturbation theory based on this Lagrangian \cite{Ecker}. 
The simplest Lagrangian of this form is
\begin{equation}
  \label{LeffLin}
  {\cal L}_{\rm eff}=\frac12 {\rm Tr} \nabla_\nu \Sigma_l \nabla_\nu
  \Sigma_l^\dagger -  \frac12 v_0 M_\pi^2 {\rm Re}
{\rm Tr}  \hat{M}_\phi \Sigma_l  -\frac12 \zeta {\rm Tr} \Sigma_l \Sigma_l^
\dagger + \frac14 \xi \left({\rm Tr} \Sigma_l \Sigma_l^\dagger\right)^2.
\end{equation}
We have used the same conventions as in \cite{KSTVZ}:
\begin{eqnarray}
  \label{def}
  \nabla_\nu \Sigma_l&=&\partial_\nu \Sigma_l - \mu (B_\nu \Sigma_l+\Sigma_l
  B_\nu^T), \nonumber \\
\nabla_\nu \Sigma_l^\dagger&=&\partial_\nu \Sigma_l^\dagger + \mu (B_\nu^T
\Sigma_l^\dagger + \Sigma_l^\dagger B_\nu), \nonumber \\
B_\nu&=&\delta_{0\nu} \left(\begin{array}{cc} 1 & 0 \\ 0 & -1
  \end{array}\right),  \\ 
\hat{M}_\phi&=&\left(\begin{array}{cc} i \sin\phi& \cos\phi \\
    \cos\phi & i \sin\phi 
  \end{array}\right), \nonumber  
\end{eqnarray}
where $\tan\phi=\lambda/m$.

The field $\Sigma_l=(v+\sigma)\Sigma$ contains $v$, the minimum of
the free energy of the linear sigma model, as well as the radial and
transverse fluctuations around that minimum. The field $\sigma$ describes the
radial fluctuations around that minimum. The three 
pseudo-Goldstone modes are the
transverse fluctuations around that minimum. They are the same as in chiral
perturbation theory and are contained in the field 
$\Sigma$ given by:
\begin{eqnarray}
\label{eqn:Sigma}
  \Sigma=U \bar{\Sigma} U^T,
\end{eqnarray}
where
\begin{eqnarray}
  U=\exp \left( \frac{i\Pi}{\sqrt{2} v} \right)
 \hspace{.5cm} {\rm with} \hspace{.5cm}
  \Pi=\left(\begin{array}{cc} P_S & Q_R+iQ_I \\ Q_R-iQ_I & P_S
  \end{array}\right),
\end{eqnarray}
and
\begin{eqnarray}
  \bar{\Sigma}=\left(\begin{array}{cc} i \sin \alpha& \cos \alpha \\
      \cos \alpha & i \sin \alpha \end{array}\right),
\end{eqnarray}
corresponds to the minimum of the free energy. 
$M_\pi$ is the pion mass in the presence of the Majorana quark mass 
$\lambda$,
\begin{equation}
M_\pi^2 = {\sqrt{m^2+\lambda^2} \over m} m_\pi^2,
\end{equation}
where $m_\pi$ is the pion mass at $\lambda=\mu=0$. The Lagrangian is
written in such a way that the masses of the pseudo-Goldstone modes
are given by $M_\pi$ at zero chemical potential.

Under a local flavor transformation $V\in U(2)$, the different
fields transform in the following way
\begin{eqnarray}
  \Sigma & \rightarrow & V \Sigma V^T \nonumber \\
  \hat{M}_\phi & \rightarrow & V^* \hat{M}_\phi V^\dagger \\
  B_\nu & \rightarrow & V B_\nu V^\dagger -\frac1\mu V\partial_\nu
  V^\dagger. \nonumber
\end{eqnarray}

    
At $\mu=0$ and $\lambda=0$, the minimum of the free energy corresponds to
\begin{equation}
v_0=\sqrt{\frac{m_\pi^2+\zeta}{2\xi}}.
\label{eqn:v_0}
\end{equation} 
In general, the minimum of the free energy is given by minimizing
\begin{equation}
{\cal E} =
\xi v^4 -\zeta v^2 -4 \mu^2 v^2\sin^2\alpha-2 M_\pi^2 v_0 v \cos(\alpha-\phi).
\label{eqn:FE}
\end{equation}
Comparing this to the phenomenological effective potential,
\begin{equation}
{\cal E} = \frac{1}{4}\,R^4 - \frac{1}{2}\,a\,R^2
         - \frac{1}{2}\,b\,\mu^2\,R^2\,\sin^2 \alpha
         - c\,m\,R\,\cos \alpha - c\,\lambda\,R\,\sin \alpha.
\label{eqn:ls1}
\end{equation}
used in our earlier work and in the next section, we see that these are
identical under the substitutions $v=\sqrt{b/2}R/2$, $\zeta=4a/b$, 
and $\xi=16/b^2$ provided the critical value of $\mu$,
$\mu_c=m_\pi/2$. 



The minimization conditions for
the ${\cal E}$ of equation~\ref{eqn:FE} are
\begin{eqnarray}
  \label{eqn:SPlin}
 v \big(2 v \mu^2  \sin 2 \alpha-v_0 M_\pi^2 \sin
   (\alpha-\phi)\big)&=&0 \\
v_0 m_\pi^2 \cos(\alpha-\phi) +v  \big( \zeta -2 v^2 \xi +4 \mu^2 \sin^2
  \alpha   \big)&=&0. \nonumber
\end{eqnarray}

The computation of the spectrum of the linear sigma model is similar to
that for chiral perturbation theory given in the appendix.  For this
case the $\sigma$ field mixes with both the $Q$-modes, but not with the
$P_S$ mode. The secular equation
for the $\sigma$ and the two $Q$-modes is given by the setting the 
determinant of the matrix
\begin{eqnarray}
  \label{eqn:secular}
 \left[
\begin{array}{c@{\hspace{-0.4in}}c@{\hspace{-0.4in}}c}
E^2-{\bf p}^2+\zeta-6 v^2 \xi+4 \mu^2 \sin^2 \alpha & 4 \mu^2 \sin
2\alpha  & 4 \mu E \sin\alpha \\
4 \mu^2 \sin 2 \alpha & E^2 -{\bf p}^2 + \zeta -2 v^2 \xi +2 \mu^2
(1+\cos 2\alpha) &
4 \mu E \cos \alpha \\
4 \mu E \sin\alpha & 4 \mu E \cos \alpha &
E^2 -{\bf p}^2 + \zeta -2 v^2 \xi  +4 \mu^2
\end{array}
\right],
\end{eqnarray}
to zero. The dispersion relation for the $P_S$ modes is 
\begin{equation}
E^2={\bf p}^2+ \frac{v_0}{v}M_\pi^2 \cos(\alpha-\phi)
\label{eqn:P_S}
\end{equation}

In order to get the masses of the different modes, the secular equation
and the dispersion relation must be solved together with the minimization
equations. The secular equation can be cast in a form more similar to that
presented in \cite{KSTVZ}, by making use of the relation
\begin{equation}
\zeta-2v^2\xi+4\mu^2=-{v \over v_0}M_\pi^2{\sin\phi \over \sin\alpha},
\end{equation}
which is a consequence of equations~\ref{eqn:SPlin}. With this form one can see
explicitly that in the ordered phase for $\lambda=0$, where $\phi=0$ but
$\sin\alpha \ne 0$, there is a true Goldstone mode.

\section{Lattice simulations and scaling}

We have simulated lattice 2-colour QCD with 1 staggered quark field (4
continuum flavours) at finite quark-number chemical potential $\mu$ on $8^4$
and $12^3 \times 24$ lattices at $\beta=4/g^2=1.5$ (close to the $\beta_c$ for
$N_t=4$) and quark mass $m=0.025$ in lattice units. Simulations were performed
with the explicit symmetry-breaking parameter, $\lambda=0.0025, \, 0.005$ (and
zero, for small $\mu$). We used the hybrid molecular-dynamics algorithm,
performing simulations of $2000$ molecular-dynamics time units at each $\mu$
and $\lambda$ with $dt$ as small as $0.0016$. The chiral and diquark
condensates, the quark number density, and the spectrum of candidate
pseudo-Goldstone bosons were measured.

\begin{figure}[htb]
\epsfxsize=4in
\centerline{\epsffile{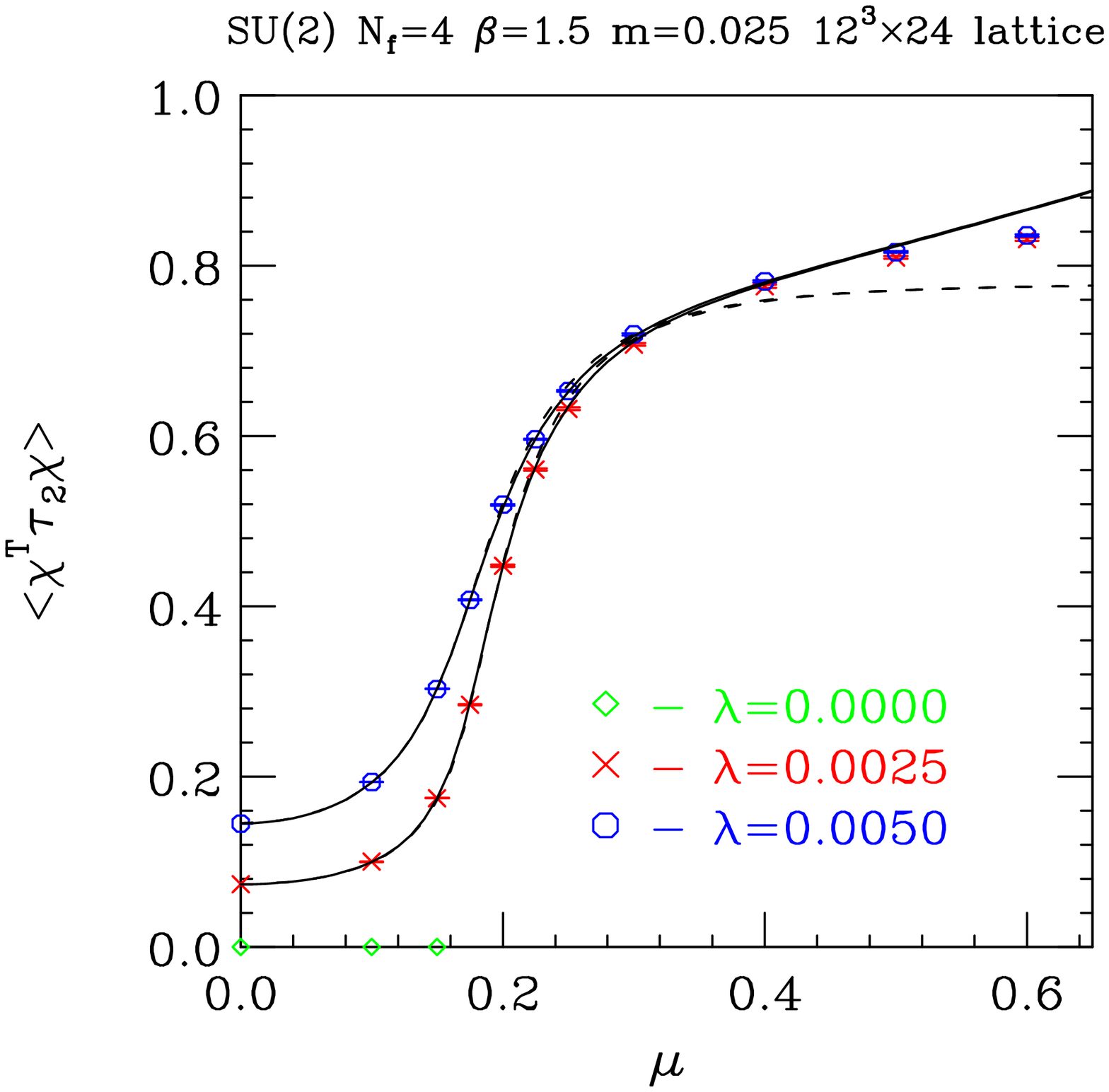}}
\centerline{\epsffile{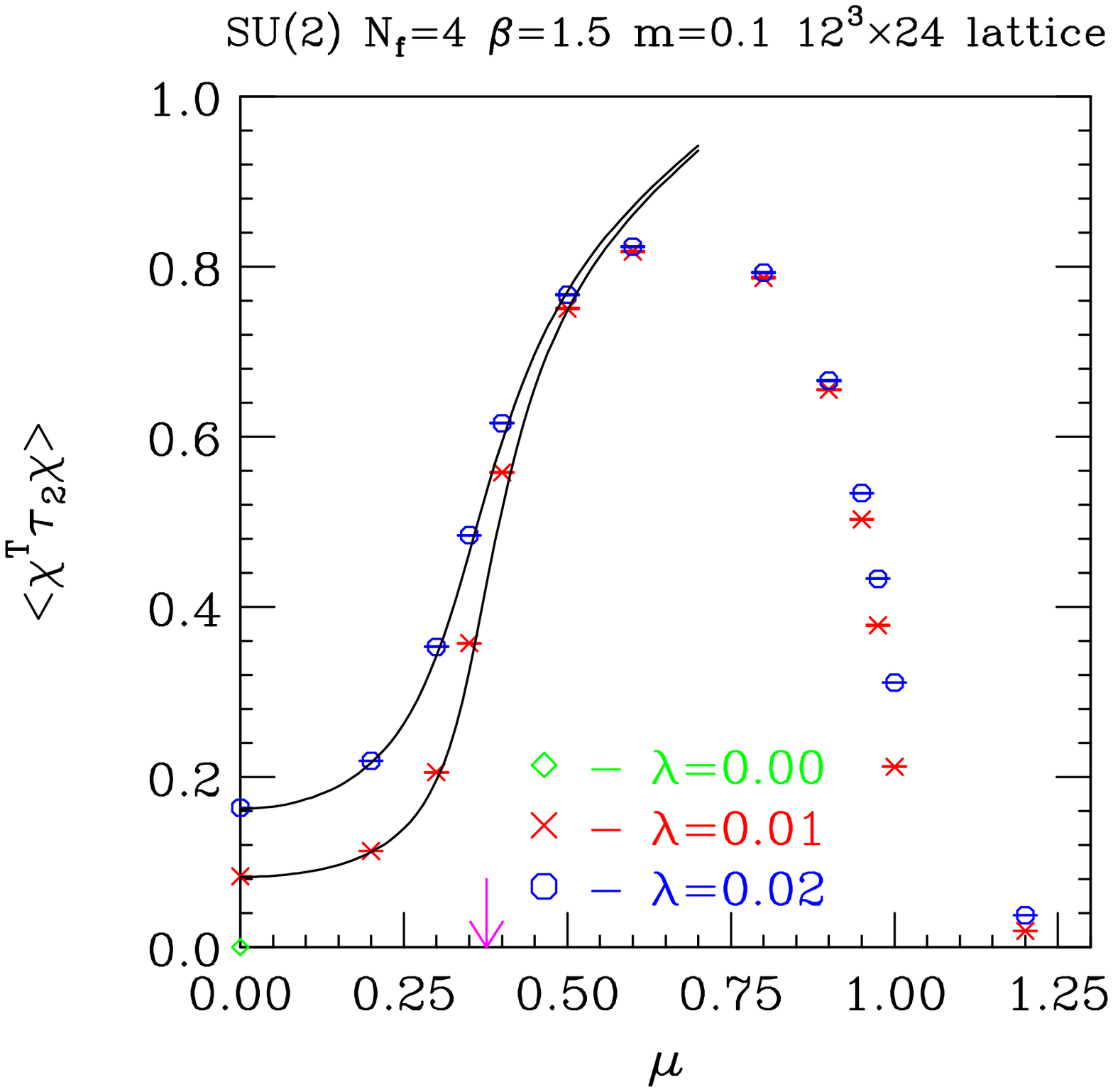}}
\caption{The diquark condensate as a function of $\mu$ on a $12^3
  \times 24$ lattice. The solid lines are fits to the form of
  equation~\ref{eqn:ls2}. The dashed lines are fits to the chiral
  perturbation theory forms in the appendix. a) $m=0.025$, b) $m=0.1$.}
\label{fig:pt2p}
\end{figure}

The diquark condensate,
$\frac{1}{2}\langle\chi^T\tau_2\chi + \bar{\chi}\tau_2\bar{\chi}^T\rangle$,
is plotted for the larger lattice in figure~\ref{fig:pt2p}a. This condensate
is seen to be small for $\mu \lesssim m_\pi/2 = 0.19264(7)$, and rapidly
increases close to $m_\pi/2$. In addition, the decrease with decreasing
$\lambda$ suggests that the condensate would vanish as $\lambda \rightarrow 0$
for $\mu \le 0.15$, while for $\mu > 0.225$, the condensate appears destined to
remain finite in this limit. These observations strongly suggest that there
is a phase transition, somewhere in the range $0.15 < \mu < 0.225$. To quantify
this observation, we have fitted the behaviour of these condensates to scaling
forms suggested from effective chiral Lagrangians. The fits to the simplest
form which comes from the tree level analysis of the chiral Lagrangian of the
non-linear sigma model variety described in the appendix are poor -- 
$\chi^2/d.o.f = 64$ --- so we turn again to a form based on the tree level
analysis of a Lagrangian of the linear sigma model class, as described in
section~3, which allows the magnitude of the condensate to vary. We have had
good experience with such a form in the past for quenched theories and for QCD
at finite isospin chemical potential \cite{quenched,isospin}.
The diquark condensate is fitted to the form
\begin{equation}
\frac{1}{2}\langle\chi^T\tau_2\chi+\bar{\chi}\tau_2\bar{\chi}^T\rangle 
               = c \: R \: \sin\alpha,
\label{eqn:ls2}
\end{equation}
which derives from equation~\ref{eqn:ls1} where $R$ and $\alpha$ are given from
the minimization conditions of the previous section. For later reference we
note that the prediction for the chiral condensate is
\begin{equation}
\langle\bar{\chi}\chi\rangle = c \: R \: \cos\alpha
\label{eqn:ls3}
\end{equation}
while the quark-number density is
\begin{equation}
j_0 = b \: \mu \: R^2 \sin^2\alpha.
\label{eqn:ls4}
\end{equation}
The constant $c$ is defined in terms of the critical $\mu$, $\mu_c$, through
\begin{equation}
c = { b \: \mu_c^2 \over m} \sqrt{a + b \: \mu_c^2},
\end{equation}
which is equivalent to equation~\ref{eqn:v_0}, provided $\mu_c=m_\pi/2$.

Our best fit to the form of equation~\ref{eqn:ls2} has $a=0.662(14)$,
$b=0.736(13)$, $m=0.02528(4)$, $\mu_c=0.19299(9)$ and $\chi^2/d.o.f.=5.4$, over
the range $0 \le \mu \le 0.4$. $\mu_c$ is in good agreement with the measured
value of $m_\pi/2 = 0.19264(7)$, while $m$ is close to the value $0.025$ used
in our simulations, considering the quality of the fit. Since the $\chi^2$ for
this fit appears poor, we compare the fit with the measurements both on the
$12^3 \times 24$ lattice which we used for the fit, and for the $8^4$ lattice
in tables~\ref{tab:pt2p0.0025},\ref{tab:pt2p0.005}.
\begin{table}[htb]
\begin{tabular}{|c|c|c|c|}
\hline
      &           \multicolumn{3}{c|}{$\lambda=0.0025$}                     \\
      &        $8^4$         &    $12^3 \times 4$    &         fit          \\
$\mu$ & $<\chi^T\tau_2\chi>$ &  $<\chi^T\tau_2\chi>$ & $<\chi^T\tau_2\chi>$ \\
\hline
0.000 &       0.0730(2)      &      0.07351(5)       &      0.07361         \\
0.100 &       0.0979(3)      &      0.1001(1)        &      0.09991         \\
0.150 &       0.1604(9)      &      0.1747(3)        &      0.17458         \\
0.175 &       0.2448(17)     &      0.2847(6)        &      0.28645         \\
0.200 &       0.3856(28)     &      0.4475(13)       &      0.44775         \\
0.225 &       0.5109(44)     &      0.5606(13)       &      0.56438         \\
0.250 &       0.5900(51)     &      0.6320(14)       &      0.63519         \\
0.300 &       0.6615(44)     &      0.7075(16)       &      0.71084         \\
0.400 &       0.7257(50)     &      0.7758(16)       &      0.77845         \\
0.500 &       0.7410(46)     &      0.8098(17)       &      0.82255         \\
0.600 &       0.7586(53)     &      0.8312(18)       &      0.86512         \\
0.800 &       0.6590(45)     &                       &                      \\
0.900 &       0.4236(39)     &                       &                      \\
1.000 &       0.0293(1)      &                       &                      \\
\hline
\end{tabular}
\caption{The diquark condensate
$\frac{1}{2}\langle\chi^T\tau_2\chi+\bar{\chi}\tau_2\bar{\chi}^T\rangle$ as a 
function of $\mu$, for $\lambda=0.0025$}\label{tab:pt2p0.0025}
\end{table}
\begin{table}[htb]
\begin{tabular}{|c|c|c|c|}
\hline
      &           \multicolumn{3}{c|}{$\lambda=0.005$}                      \\
      &        $8^4$         &    $12^3 \times 4$    &         fit          \\
$\mu$ & $<\chi^T\tau_2\chi>$ &  $<\chi^T\tau_2\chi>$ & $<\chi^T\tau_2\chi>$ \\
\hline
0.000 &       0.1437(3)      &      0.1451(1)        &      0.14518         \\
0.100 &       0.1902(5)      &      0.1936(2)        &      0.19316         \\
0.150 &       0.2893(12)     &      0.3030(4)        &      0.30342         \\
0.175 &       0.3826(19)     &      0.4077(8)        &      0.40704         \\
0.200 &       0.4883(23)     &      0.5191(8)        &      0.51529         \\
0.225 &       0.5820(29)     &      0.5962(9)        &      0.59698         \\
0.250 &       0.6275(33)     &      0.6529(10)       &      0.65228         \\
0.300 &       0.6963(33)     &      0.7193(12)       &      0.71715         \\
0.400 &       0.7509(37)     &      0.7817(11)       &      0.78057         \\
0.500 &       0.7795(38)     &      0.8162(12)       &      0.82395         \\
0.600 &       0.7969(40)     &      0.8354(12)       &      0.86628         \\
0.800 &       0.7078(33)     &                       &                      \\
0.900 &       0.4945(27)     &                       &                      \\
1.000 &       0.0584(2)      &                       &                      \\
\hline
\end{tabular}
\caption{The diquark condensate 
$\frac{1}{2}\langle\chi^T\tau_2\chi+\bar{\chi}\tau_2\bar{\chi}^T\rangle$ as a 
function of $\mu$ for $\lambda=0.005$}\label{tab:pt2p0.005} 
\end{table}
While the difference between the $8^4$ and $12^3 \times 24$ `data' suggest that
the finite size effects in the measurements on the larger lattice are small,
they are almost certainly comparable with and probably larger than the
statistical errors and the discrepancies between the `data' and the fits given
in these tables, over the range of the fits. For this reason we consider the
fits to the linear sigma model form to be acceptable, indicating that the
system undergoes a second order transition with mean-field critical exponents
at $\mu=m_\pi/2$. We have also shown the fit to the tree-level chiral 
perturbation theory mentioned above in our figure. While this clearly
has a more limited range of validity, it does not appear that unreasonable over
this range. However, in order to obtain this quality of fit, we were forced to
use $m=0.02632(2)$. This is far enough from the true mass $m=0.025$ that, as
we shall see, the prediction for the chiral condensate is considerably poorer. 



Since this fit gives not only the $\mu$ and $\lambda$
dependence but also predicts the $m$ dependence, we have plotted the
predictions of this fit for our old `data' at $m=0.1$ \cite{HKMS} in
figure~\ref{fig:pt2p}b. Considering the fact that $m=0.1$ is rather large to
expect fits aimed at the chiral limit to work well, the prediction is
remarkably good below $\mu=0.6$ where the effects of saturation start to be
seen.

On the smaller lattice we see that the diquark condensate has a broad peak 
near $\mu=0.6$, beyond which it falls, remaining very small above 
$\mu \approx 1$. Since, as we shall see later, the quark-number density 
approaches 2, the maximum value allowed by fermi statistics, at these $\mu$
values, we interpret this fall as a saturation effect, a finite lattice spacing
artifact. Further evidence that this is indeed a lattice artifact is found by
comparing simulations at different lattice spacings. Comparing the results
presented here at $\beta=1.5$ with those at $\beta=1.85$ \cite{KTS2}
where the lattice spacing is about $2/3$ that at $\beta=1.5$, we notice that
the value of $\mu$ in lattice units where saturation is reached is consistent
with being the same in both cases. In addition, the $\mu$ of the peak in the
diquark condensate for $\beta=1.85$ (in lattice units) is at least as large as
at $\beta=1.5$. If these were continuum effects the relevant lattice $\mu$
values for $\beta=1.85$ would be smaller by roughly a factor of $2/3$, since
$\mu = \mu_{physical} a$. Hence we conclude that the decrease in
$\frac{1}{2}\langle\chi^T\tau_2\chi+\bar{\chi}\tau_2\bar{\chi}^T\rangle$ at
large $\mu$ is purely a lattice artifact.
Tables~\ref{tab:pt2p0.0025},\ref{tab:pt2p0.005} indicate that finite size
effects increase with $\mu$, and that their effect is to depress the values of
the condensate as $\mu$ increases. Thus we should expect that the infinite
lattice peak will be at a higher $\mu$ value. However, it is probable that
this value will still lie below the saturation $\mu$, so that the falloff will
occur over a range of $\mu$, even on an infinite lattice, rather than as an
abrupt discontinuity at the saturation value of $\mu$. We come to this
conclusion based on our experience with the quenched theory, where we have
been able to examine the finite size effects more thoroughly \cite{quenched}.

\begin{figure}[htb]                                                        
\epsfxsize=4in                                                          
\centerline{\epsffile{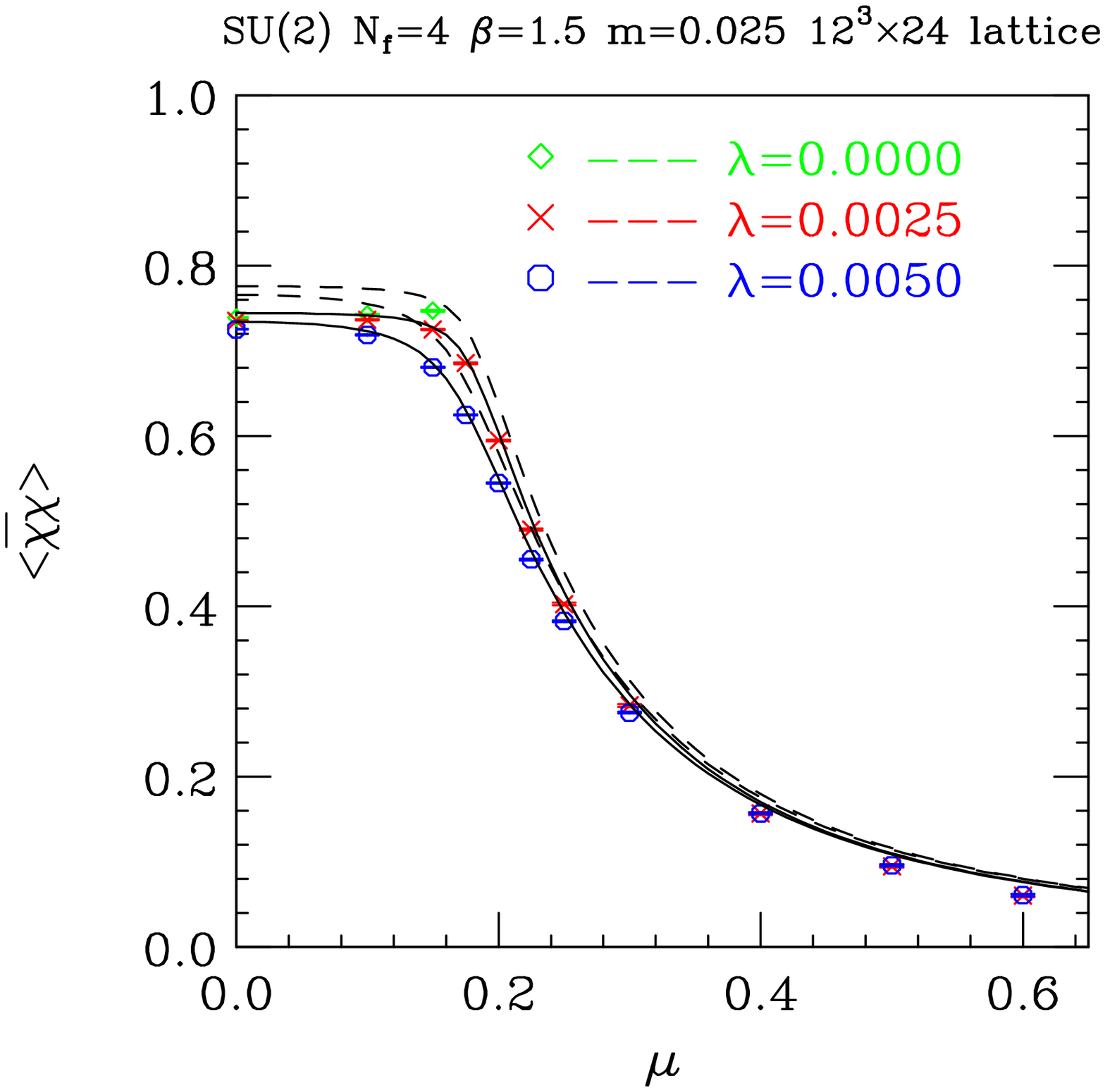}}                                        
\centerline{\epsffile{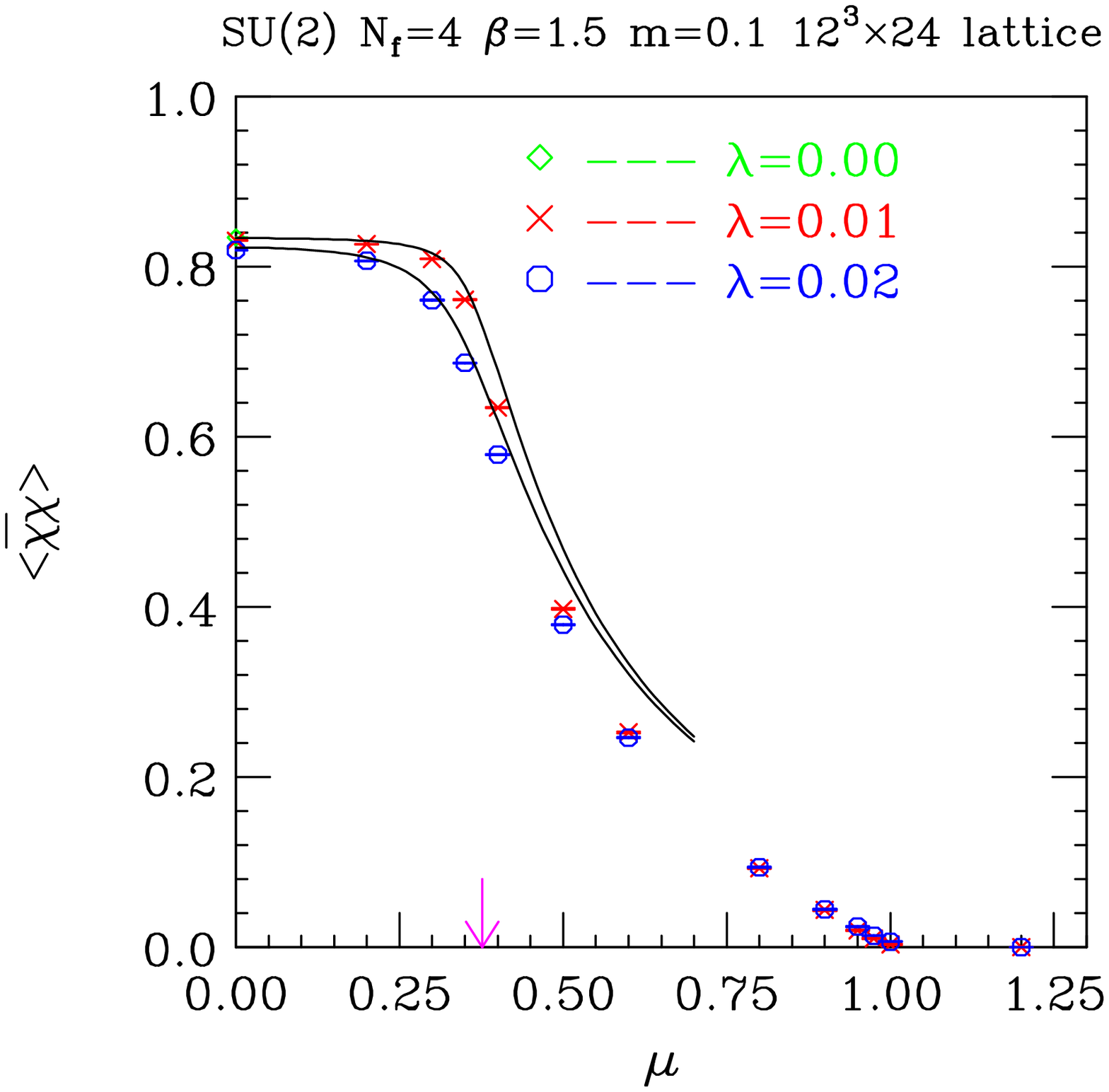}}                                        
\caption{The chiral condensate as a function of $\mu$ on a $12^3 \times 24$  
         lattice. The solid lines are from the fits to the diquark condensate.
         The dashed lines are the predictions of the fit to the tree-level
         chiral perturbation theory form. a) $m=0.025$, b) $m=0.1$. 
}
\label{fig:pbp}
\end{figure}

In figure~\ref{fig:pbp}a we show the chiral condensate,
$\langle\bar{\chi}\chi\rangle$, as a function of $\mu$. As expected it remains
approximately constant for $\mu < m_\pi$, above which it falls towards zero.
The predictions of equation~\ref{eqn:ls3}, using the parameters obtained from
fits to the diquark condensates, are plotted on this graph. The agreement
appears excellent over the scaling window. For comparison we include the
predictions for $m=0.1$ on our old `data' at that mass. Here the agreement is
considerably poorer. Presumably some of this is due to $m=0.1$ being too high
for these chiral effective Lagrangians. In addition, the chiral condensate
is expected to be more sensitive to cutoff effects, such as saturation, at
higher quark mass. Note that the prediction from the chiral perturbation theory
fit mentioned above does relatively poorly, because this predicts that the
magnitude of the total condensate is independent of $\mu$ which is not true.
Thus by forcing this form to fit the diquark condensate, the prediction for
the chiral condensate must fail.

\begin{figure}[htb]                                                            
\epsfxsize=4in                                                                
\centerline{\epsffile{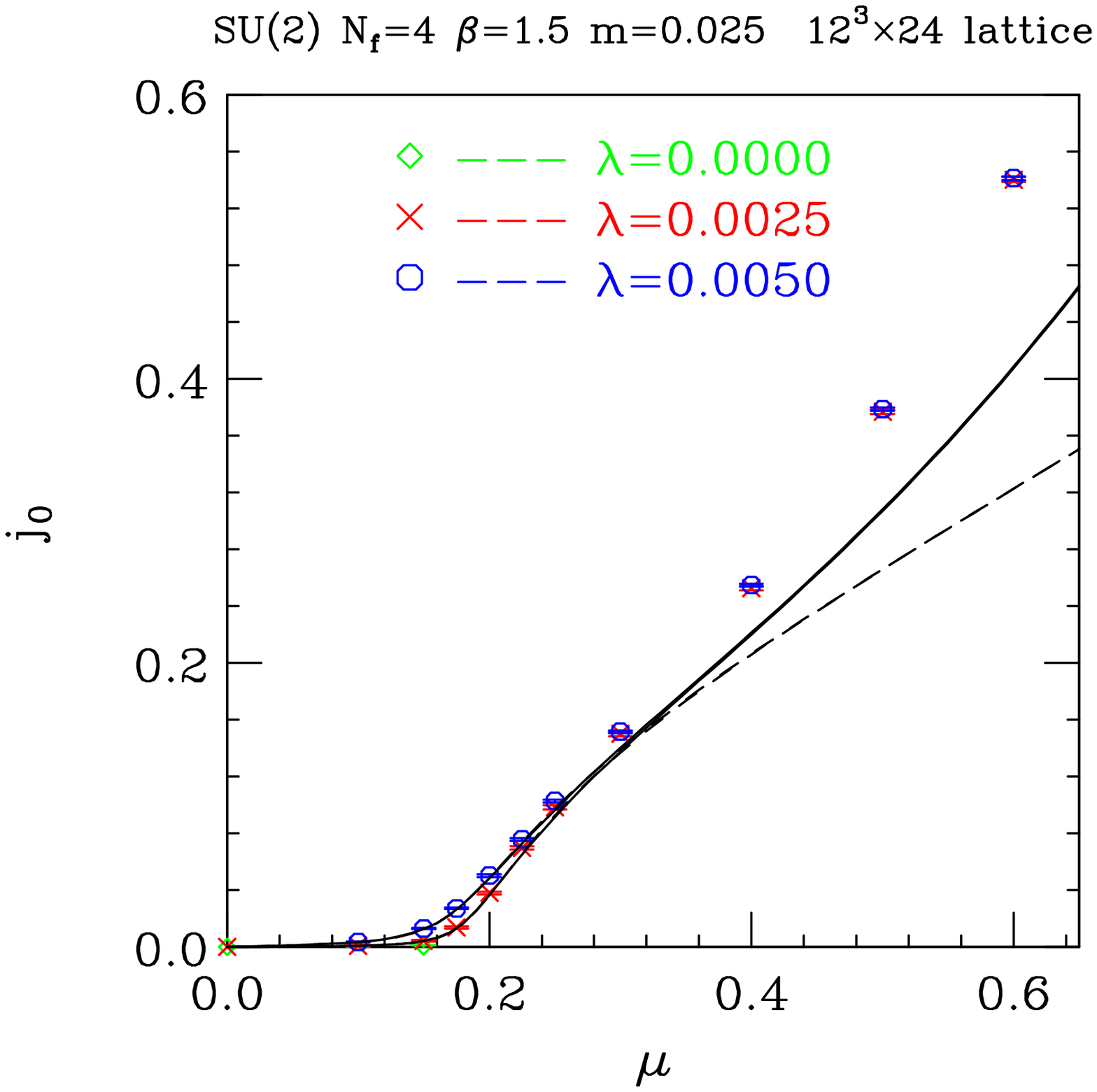}}                         
\centerline{\epsffile{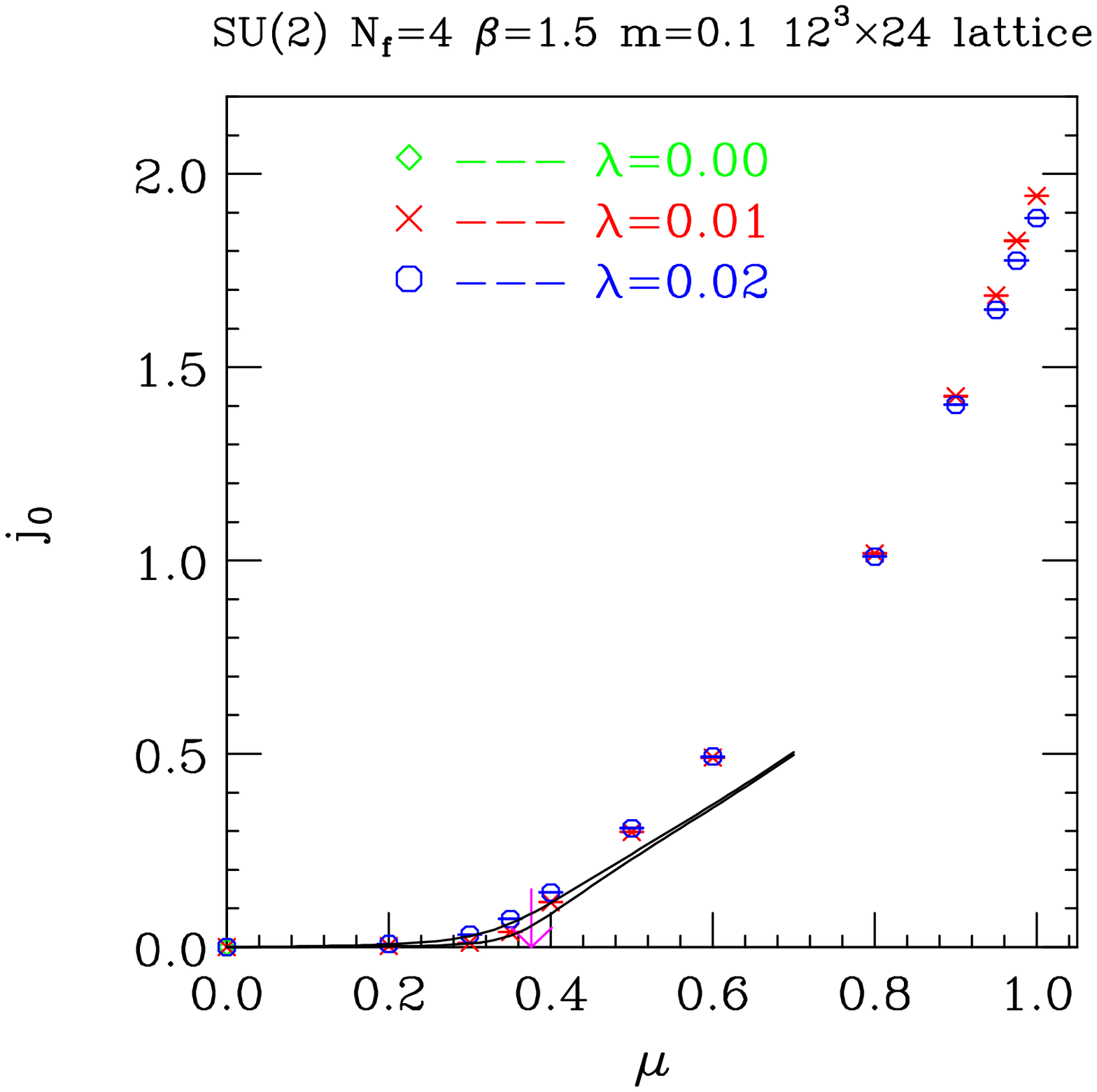}}              
\caption{The quark-number density $j_0$ as a function of $\mu$ at $\beta=1.5$,
         $m=0.025$ on a $12^3 \times 24$ lattice. The solid curves are the 
         predictions from equation~\ref{eqn:ls4}. The dashed curves are from
         tree-level chiral perturbation theory. 
         a) $m=0.0025$ b) $m=0.1$.
}
\label{fig:j0_12}
\end{figure}

\begin{figure}[htb]                                                             
\epsfxsize=4in                                                                  
\centerline{\epsffile{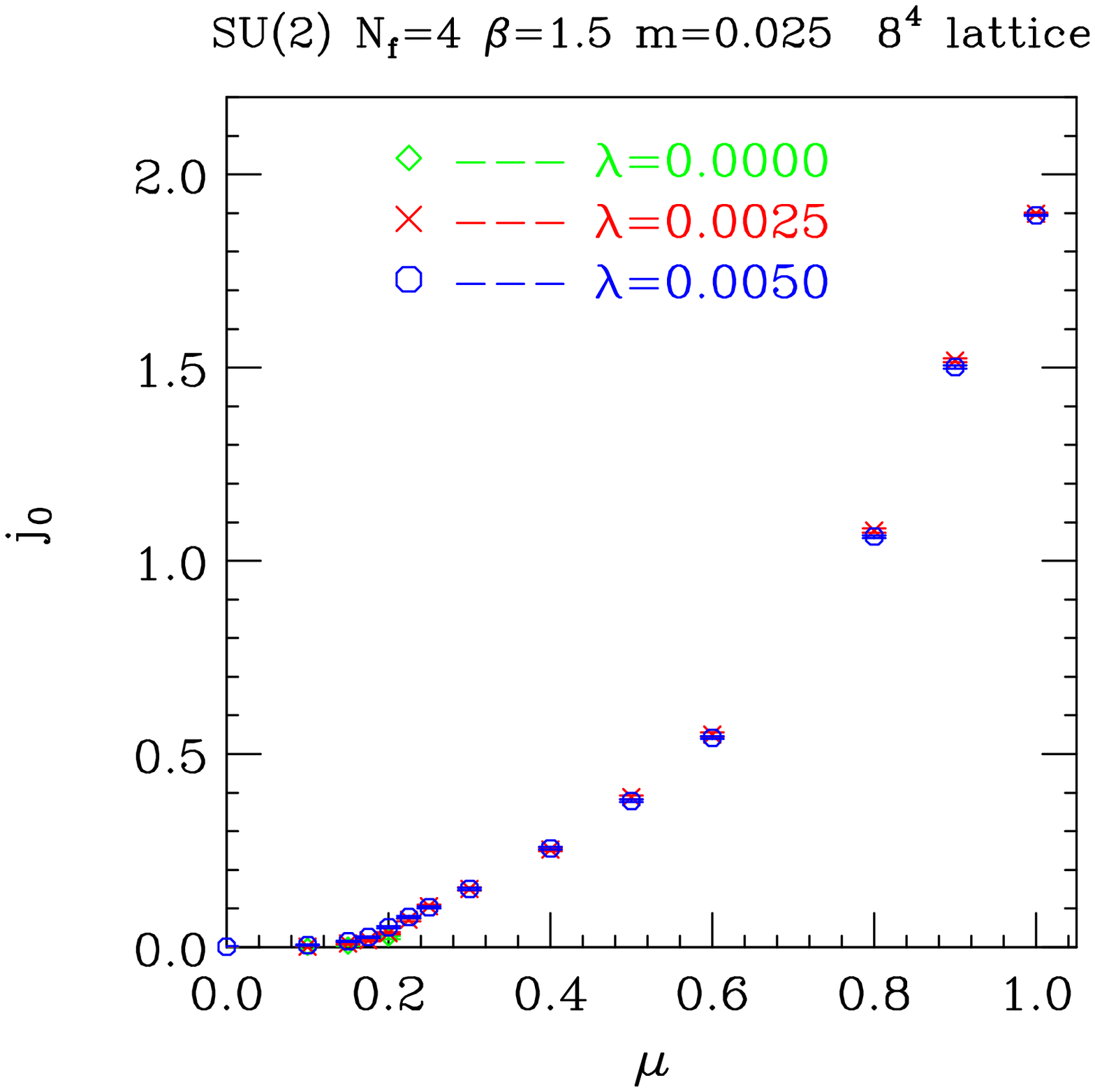}}
\centerline{\epsffile{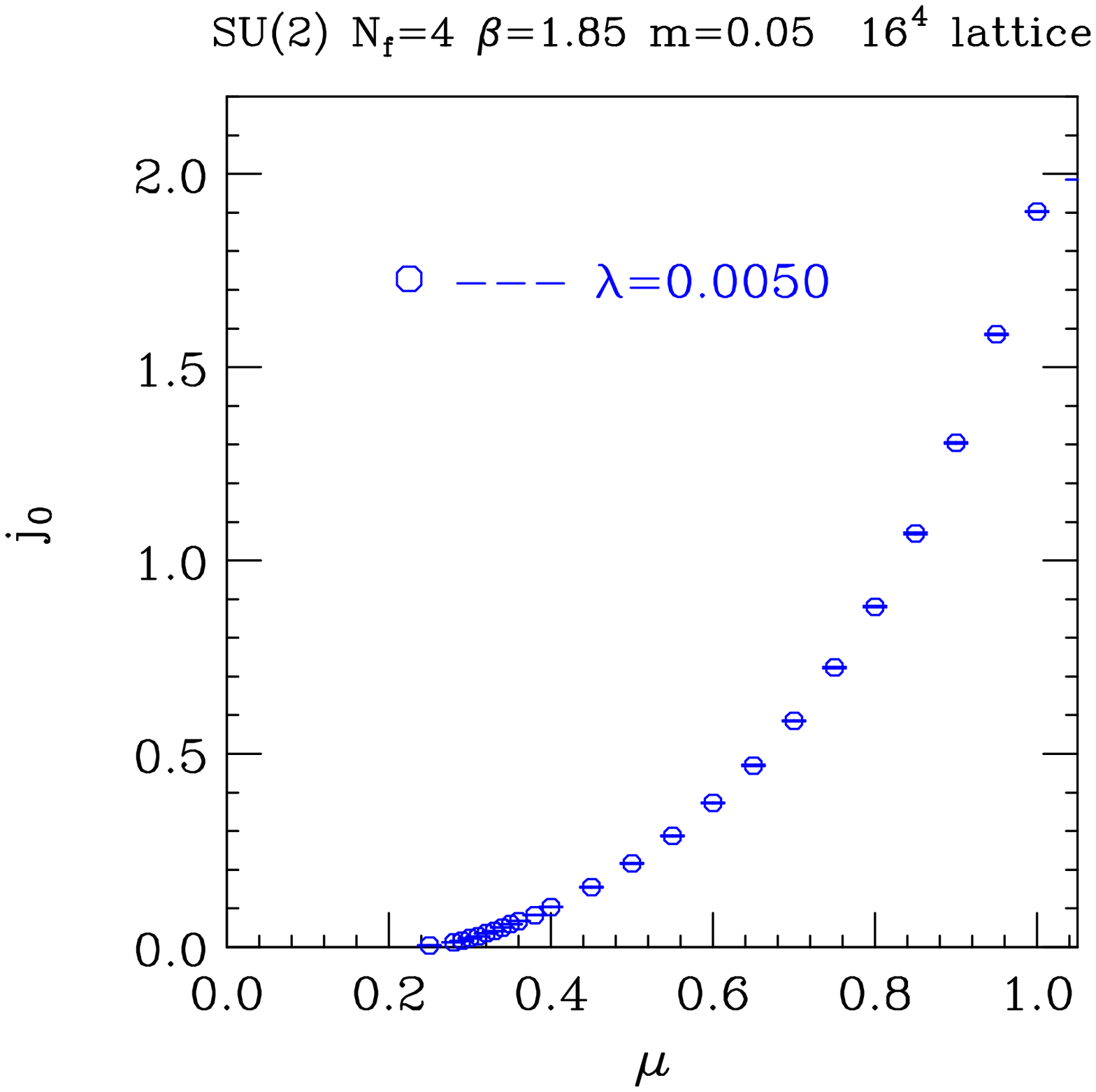}}                                          
\caption{The quark-number density $j_0$ as a function of $\mu$ at $\beta=1.5$,  
         $m=0.025$ on a $12^3 \times 24$ lattice. The curves are the predictions
         from equation~\ref{eqn:ls4}.}
\label{fig:j0}                                                               
\end{figure}

Finally we present the quark number density $j_0$ as a function of lattice $\mu$
in figure~\ref{fig:j0_12}a for our $12^3 \times 24$ simulations. Since this data
does not extend to the saturation region, we also plot the $8^4$ `data' in
figure~\ref{fig:j0}, along with the $\beta=1.85$ results for comparison. Our
corresponding `data' from our old $m=0.1$ runs is plotted, along with the
predictions from equation~\ref{eqn:ls4}, in figure~\ref{fig:j0_12}b. 

First we note that the $12^3 \times 24$ measurements start to depart from the
predictions from our scaling form (equation~\ref{eqn:ls4}) somewhat earlier than
do either of the condensates, namely for $\mu > 0.3$. For $m=0.025$, this does
admit a small scaling window. For $m=0.01$, since $0.3 < m_\pi/2$, there is
no scaling window. The fact that the relevant variable appears to be $\mu$
rather than $2\mu/m_\pi$, suggests that at least some of this departure
is a lattice artifact related to saturation.
 
As is observed in figure~\ref{fig:j0}a, the quark-number density rises even
more rapidly at larger $\mu$, until it saturates at $2$ near $\mu=1$.
Comparing these $\beta=1.5$ measurements with those at $\beta=1.85$
(figure~\ref{fig:j0}b) indicates that saturation occurs at approximately the
same $\mu$ in lattice units, confirming that saturation is a lattice artifact.
However, $j_0$ would reach saturation at the same lattice $\mu$ independent of
lattice spacing, if $j_0 \propto \mu^3$, at large $\mu$. This is precisely the
behaviour expected at large $\mu$ in the continuum. (This was pointed out by
Son and Stephanov for QCD at finite isospin density \cite{sonstep}.) The
linear rise in the diquark condensate at large $\mu$, which is a property of the
linear-sigma-model fitting forms does predict a cubic rise in $j_0$, but in
every case we have considered the measured $j_0$ far exceeds our predictions.
The hint that this might be real, i.e. not completely attributable to
saturation, comes from the observation that the onset of this rapid rise in
$j_0$ with $\mu$ appears to occur at a larger value of $\mu$ in lattice units
at $\beta=1.5$ (figures~\ref{fig:j0_12}b,\ref{fig:j0}a) than at $\beta=1.85$
(figure~\ref{fig:j0}b). This is what would be expected if it is a real effect
rather than a lattice artifact. This earlier departure from the predictions
from our tree-level analysis of our effective Lagrangian should not come as
too much of a surprise. This analysis parametrizes the departure from 
tree-level chiral perturbation theory by a single new parameter. This parameter
is set by our fit to the diquark condensate. Our new Lagrangian retains
the same relationship between the diquark condensate and the density as the
chiral perturbation theory Lagrangian. What we are seeing is evidence that
this relationship breaks down at a $\mu$ value considerably less than the
saturation value. To retain agreement beyond this point would require more
terms/parameters in our effective Lagrangian.



\section{The pseudo-Goldstone spectrum}

As discussed in section~2 and made quantitative in section~3, spontaneous
breaking of the lattice $U(2)$ flavour symmetry at $m=\mu=\lambda=0$ should
give rise to 3 Goldstone bosons. When these parameters are small, but non-zero,
these excitations become pseudo-Goldstone bosons, gaining masses dependent on 
the magnitude of these symmetry-breaking parameters. We have measured the
connected and disconnected contributions to the propagators of all local 
scalar and pseudoscalar diquarks in our $12^3 \times 24$ runs every 1 
molecular-dynamics time step. The connected propagators are measured using
noisy estimators of a point source on each odd respectively even site/colour
of 1 time-slice of the lattice. The disconnected propagators are calculated
using 5 sets of noisy sources defined over the whole odd respectively even
sublattice, and the noise-diagonal terms are discarded.

The first state considered is the scalar diquark created by applying the
operator $\frac{1}{2}(\chi^T\tau_2\chi - \bar{\chi}\tau_2\bar{\chi}^T)$ to the
vacuum. For $\lambda=0$ this will be a true Goldstone boson in the diquark
condensed phase. With the small $\lambda$s we consider, it should have a small
mass in the broken phase. Tree-level analysis of the chiral perturbation
theory Lagrangian discussed in the appendix predicts that, at finite $\lambda$
its mass should be given by equation~\ref{eqn:dispRel} for the state labeled
$\tilde{Q}$. Note that at $\lambda=0$, this reproduces the predictions of
simpler arguments $m_G=m_\pi-2\mu$ for $\mu < m_\pi/2$ and zero for $\mu >
m_\pi/2$. For the linear sigma model approach of section~3 this is replaced by
the lowest lying solution of the secular equation obtained from
equation~\ref{eqn:secular}. At $\lambda=0$ the 2 forms are identical.

\begin{figure}[htb]
\epsfxsize=4in
\centerline{\epsffile{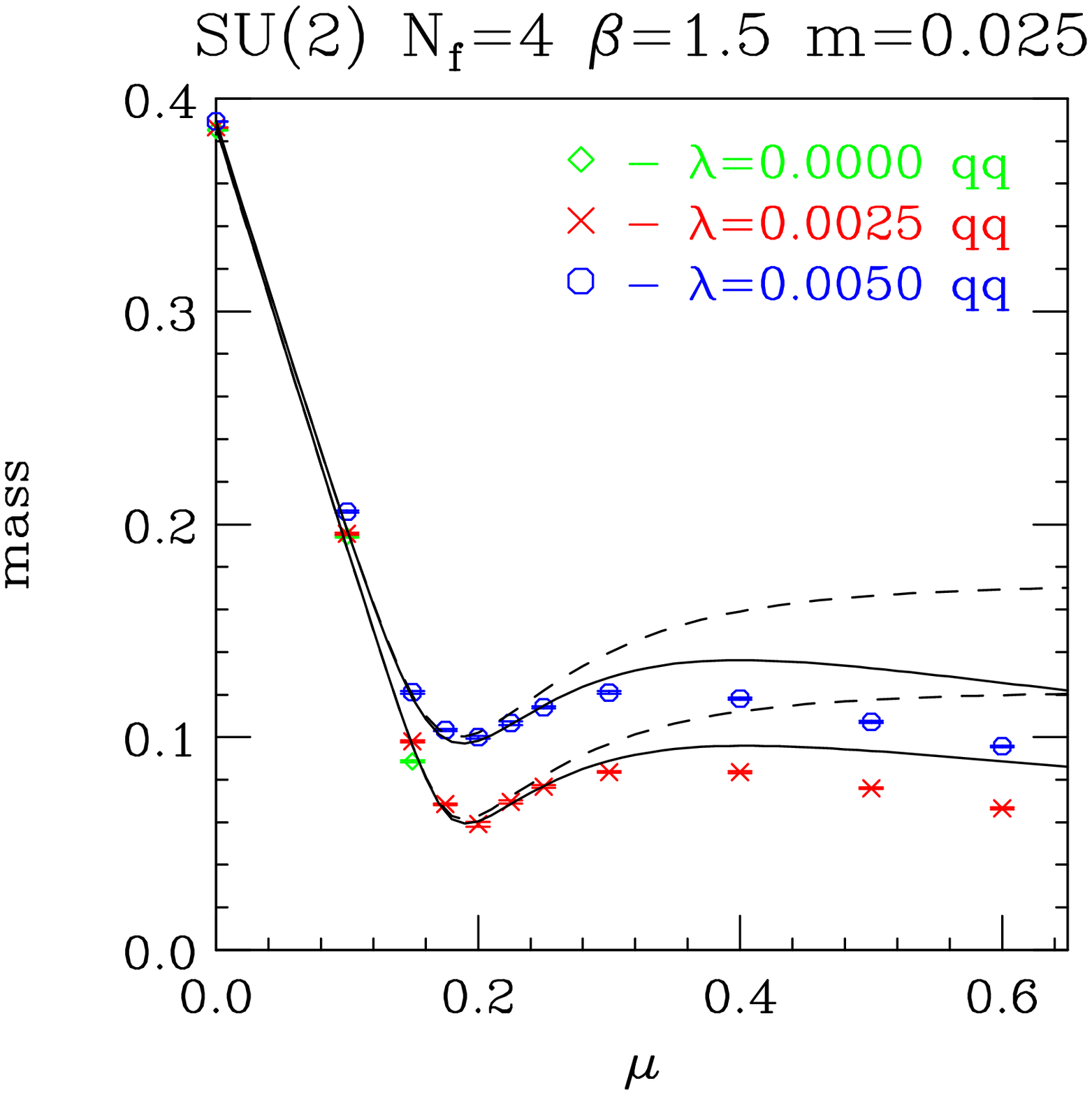}}
\centerline{\epsffile{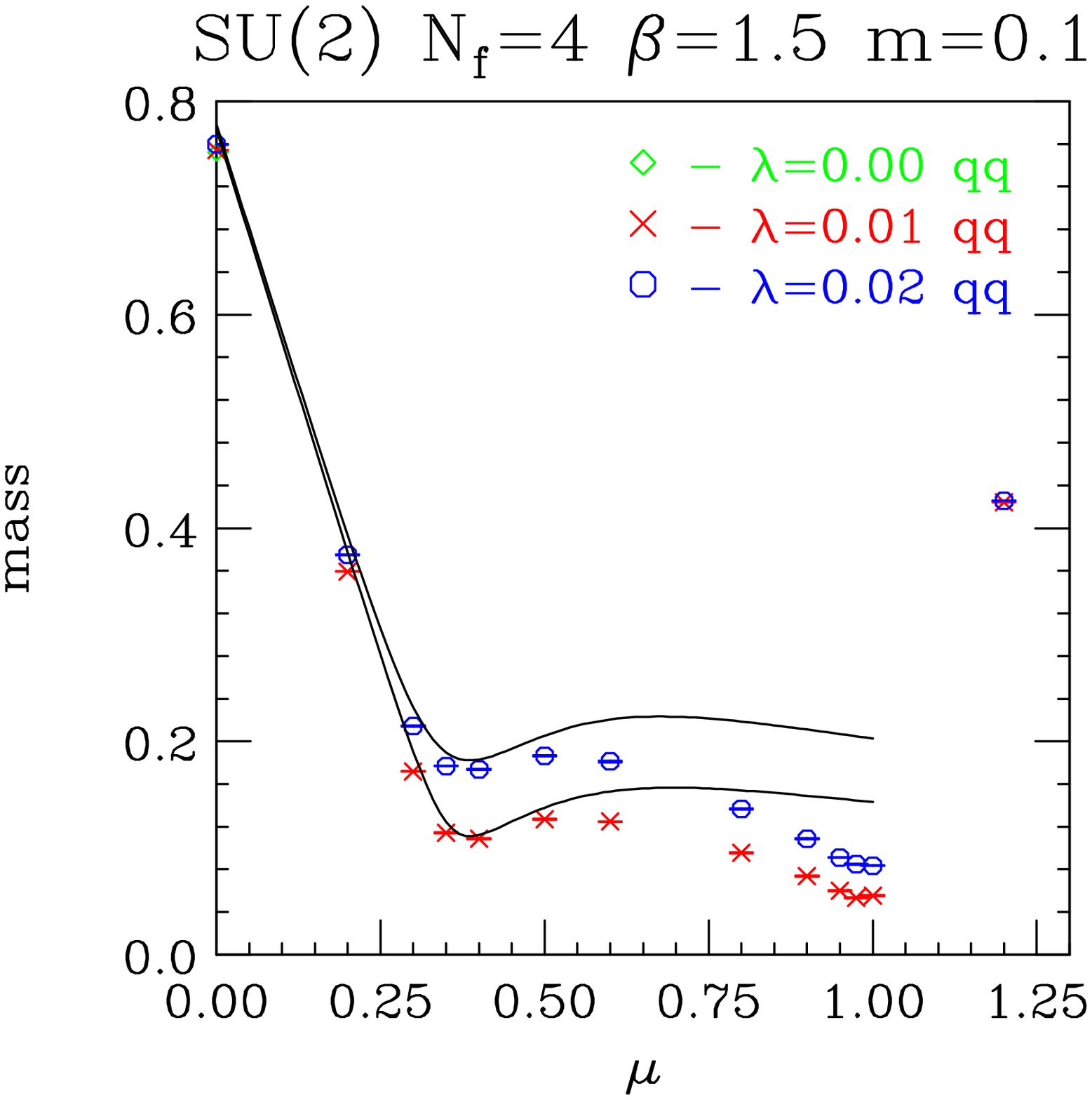}}
\caption{The Goldstone boson of quark-number violation as a function of $\mu$.
a) m=0.025, b) m=0.1. The solid curves are the predictions from our fits. The
dashed curves are from tree-level chiral perturbation theory.}
\label{fig:goldstone}
\end{figure}

We fit our measured propagators $P_G$ to the form
\begin{equation}
P_G(t) = A \{ \exp[-m_G t] + \exp[-m_G (N_t-t)] \}
\end{equation}
giving the results shown in figure~\ref{fig:goldstone}a. The solid curves in
this figure are the predictions from equation~\ref{eqn:secular}. The dashed
curves are from equation~\ref{eqn:dispRel}.
What we see is that equation~\ref{eqn:secular} describes the decrease in mass of
this would-be Goldstone boson well for $\mu < m_\pi/2$, and the dip near
the transition value $m_\pi/2$. As $\mu$ is increased much beyond $m_\pi/2$, the
`data' falls below these predictions. These linear sigma model predictions
are a considerable improvement over those of tree level chiral perturbation
theory. This gives us confidence, that even though these fits fail above the
transition, this excitation will still have the expected behaviour in the
limit $\lambda \rightarrow 0$, indicating that there is a phase transition to
a state with a diquark condensate, and that this state is the massless
Goldstone boson associated with the spontaneous breaking of quark number.
In figure~\ref{fig:goldstone}b we compare our predictions with our earlier 
measurements at $m=0.1$, and find the agreement to be somewhat worse than at
$m=0.025$.

Such departures from the predictions of the linear-sigma-model form
could well indicate that this model of the higher order terms in chiral
perturbation theory is too naive to correctly predict more than the qualitative
nature of the pseudo-Goldstone spectrum. The worse agreement for $m=0.1$,
where in terms of the scaling variable $x=2\mu/m_\pi$ saturation occurs
much sooner, suggests that the deviation could be largely due to saturation.

\begin{figure}[htb]
\epsfxsize=4in
\centerline{\epsffile{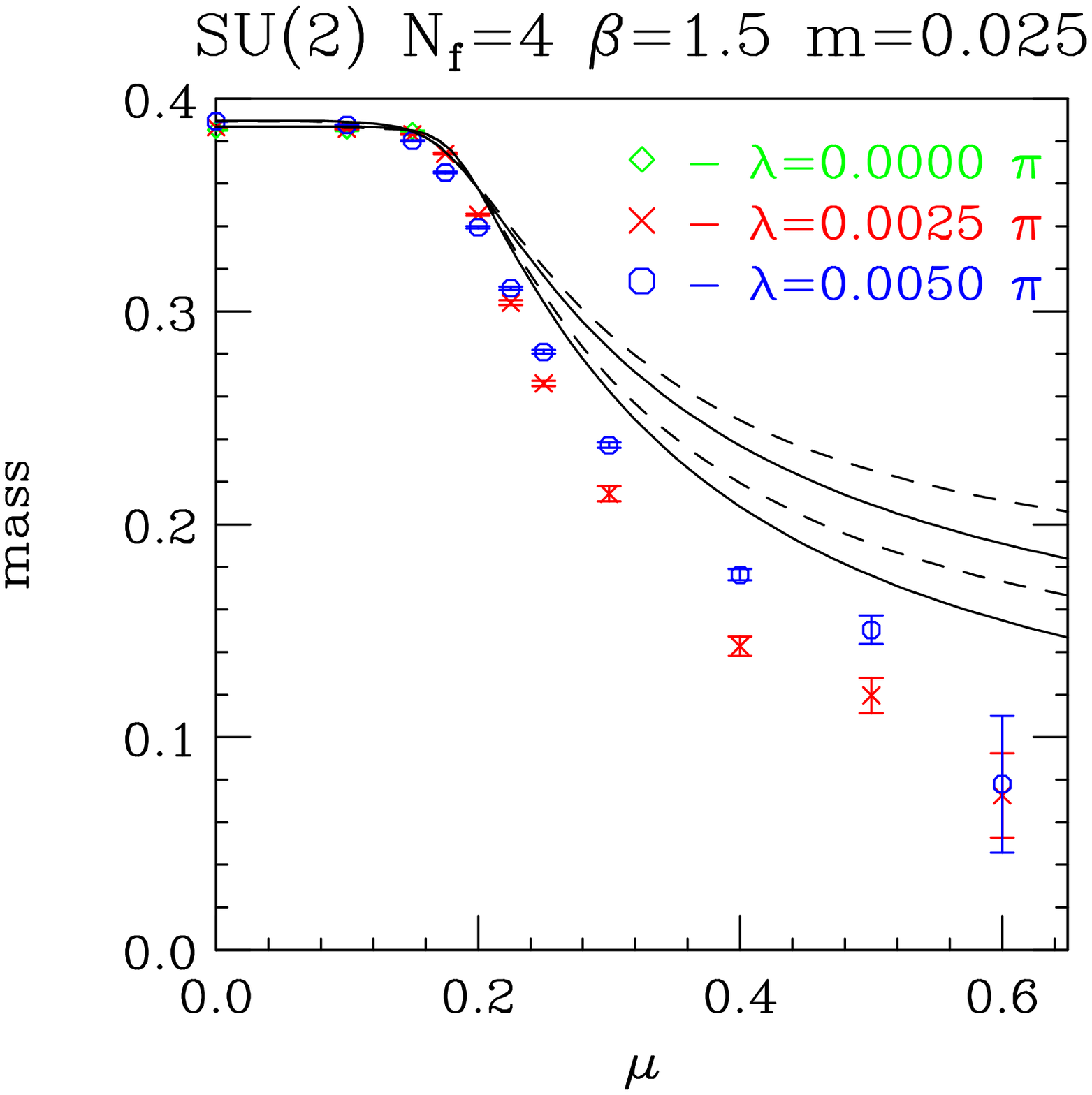}}
\centerline{\epsffile{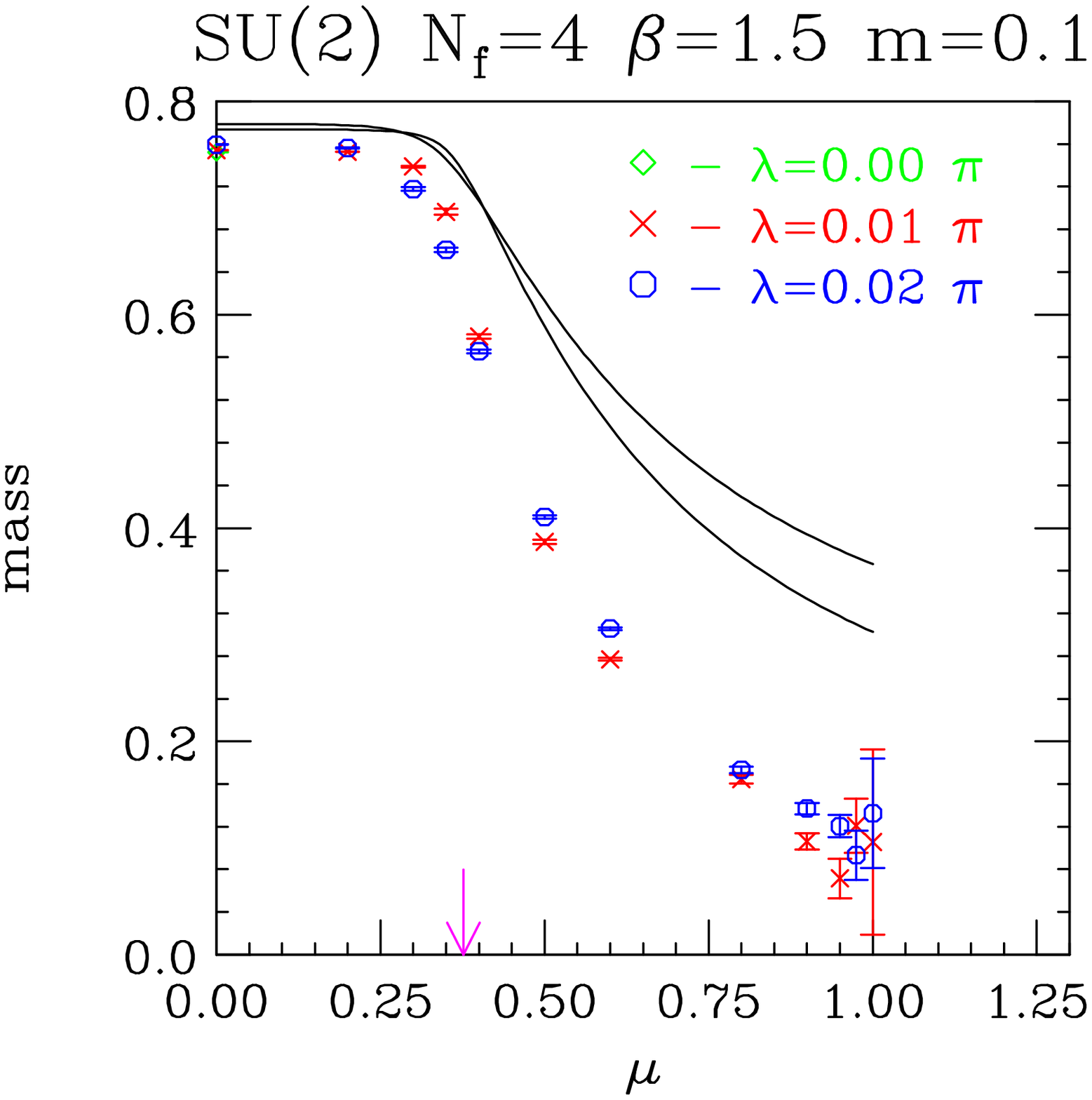}}
\caption{The pion mass as functions of $\mu$, on a $12^3 \times 24$ lattice at
a) $m=0.025$, b) $m=0.1$. The curves are the scaling predictions mentioned in 
the text.}
\label{fig:pi}
\end{figure}

\begin{figure}[htb]                                                      
\epsfxsize=4in                                                          
\centerline{\epsffile{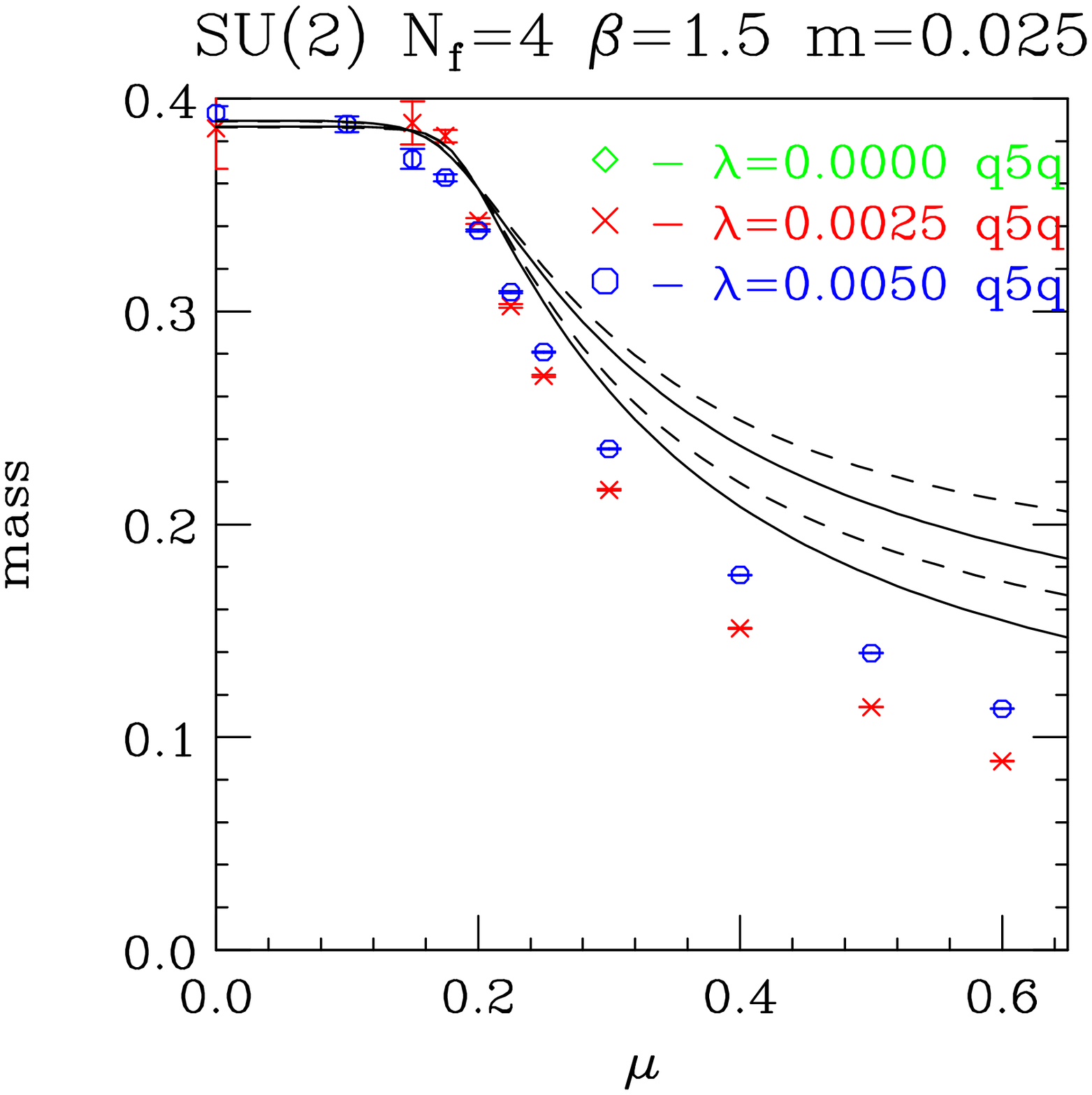}}                              
\centerline{\epsffile{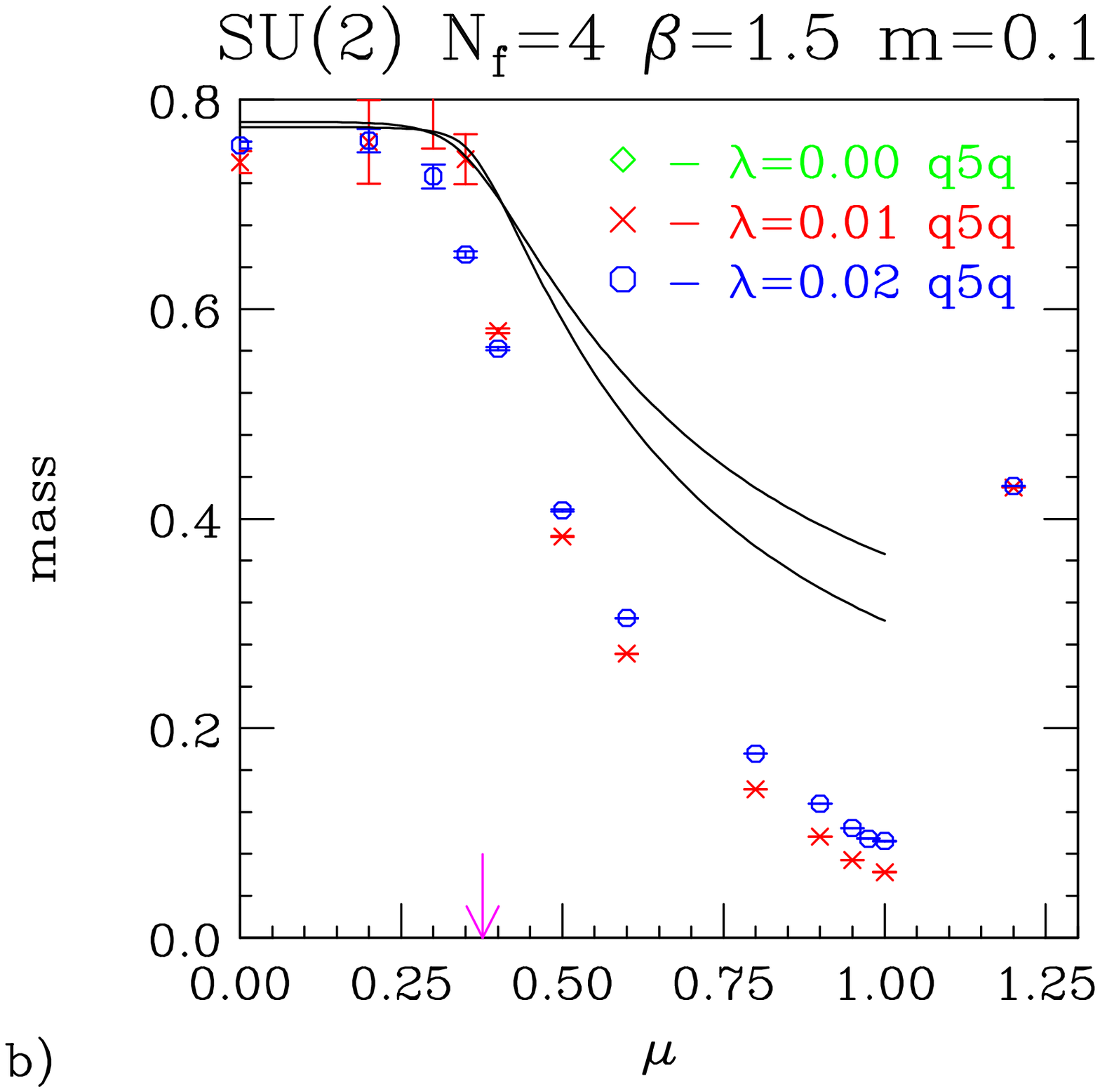}}                              
\caption{The pseudoscalar diquark mass as functions of $\mu$, on a $12^3
\times 24$ lattice at a) $m=0.025$, b)$m=0.1$. The curves are the scaling
predictions mentioned in the text.}
\label{fig:q5q}                                                                
\end{figure}

Next we turn to the consideration of the pseudoscalar pseudo-Goldstone boson. 
As described in section~2 and in particular in equation~\ref{eqn:pi}, at
$\lambda=0$ this is expected to be the pion for $\mu < m_\pi/2$. Above this
value it mixes with the pseudoscalar diquark. For large $\mu$ it should be
predominately a pseudoscalar diquark. Since the mixing angle $\bar{\alpha}$ in
equation~\ref{eqn:pi} is not known, beyond the predictions of effective
Lagrangians which suggest that $\alpha=\bar{\alpha}$, we choose to fit the
pion and pseudoscalar diquark propagators separately to a form analogous to
that which we used for the Goldstone propagator. The graphs of the measured
masses are presented in figures~\ref{fig:pi},\ref{fig:q5q}.
We note that these masses have the expected behaviour in that they remain flat
from $\mu=0$ to the vicinity of $m_\pi/2$, and then commence to fall, becoming
small for large $\mu$ where they should equal the masses of the would-be
Goldstone boson. As expected these 2 estimators for the mass of the pseudoscalar
pseudo-Goldstone boson, are in good agreement.

The effective (chiral) Lagrangian analysis presented in the appendix predicts
that this pseudoscalar boson should have a mass given by the $P_S$ state
of equation~\ref{eqn:dispRel}, while our linear-sigma model form is given in 
equation~\ref{eqn:P_S}. We have plotted these curves on our `data' in
figures~\ref{fig:pi}a,\ref{fig:q5q}a. Although these predictions have
qualitatively the same form as the `data', clearly there is no quantitative
agreement. (Note that the linear sigma model form of section~3 gives a slightly
better fit than the chiral perturbation theory form in the appendix.) The
form of the mass of equation~\ref{eqn:P_S} suggests that we compare our
data with that of our earlier paper where $m=0.1$. This we do in 
figures~\ref{fig:pi}b,\ref{fig:q5q}b. Again we find the falloff in mass above
$m_\pi/2$ is much more rapid for $m=0.1$ than for $m=0.025$. This comparison
suggests that at least some of this is the effect of saturation, but if it is
all a saturation effect, the range over which saturation has an effect is
large. Any discrepancy between the measurements and fits which remains after
removal of the effects of saturation should again be taken as an indication
that our linear sigma model effective Lagrangian is inadequate to describe all
departures from tree-level chiral perturbation theory.

We note that the term in each of our effective Lagrangians which relates
$j_0$ to the diquark condensate is the term proportional to $\mu^2$. As we
see in equation~\ref{Leffj} of the appendix and its equivalent for the 
linear sigma model, this term also contributes to the masses of the 
pseudo-Goldstone bosons. Thus deviations of $j_0$ from our predictions, should
imply differences in the pseudo-Goldstone boson masses from our predictions.
The contribution to the state which is a true Goldstone boson at $\lambda=0$,
is proportional to $\sin^2\alpha$, while that for the pseudoscalar is
proportional to $\cos^2\alpha$. Since $\alpha$ rises from zero above 
$\mu=m_\pi/2$, we would expect any departure from our predictions to occur
earlier for the pseudoscalar than for the would-be Goldstone boson, which is
precisely what we see.

\begin{figure}[htb]
\epsfxsize=6in
\centerline{\epsffile{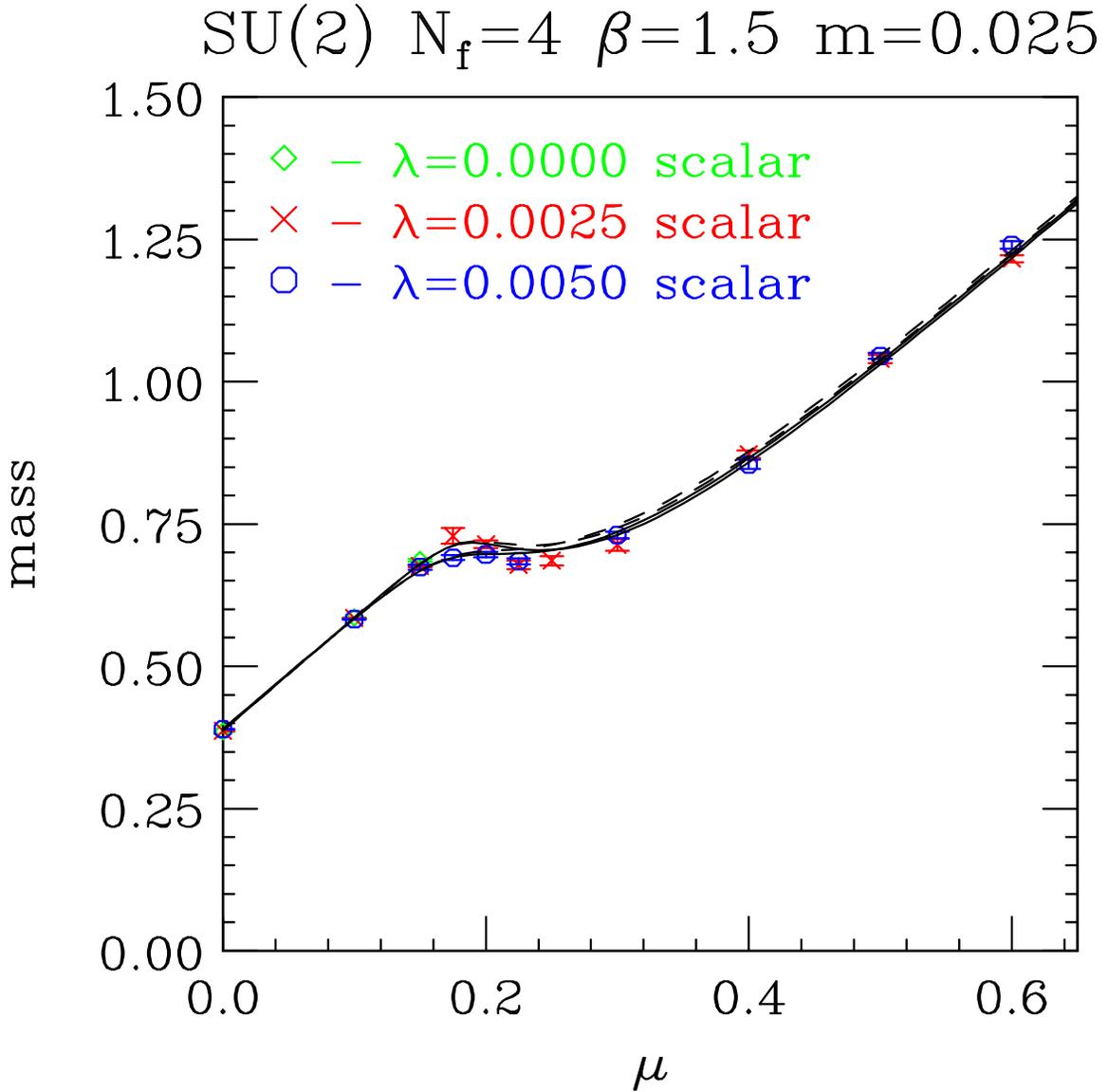}}
\caption{The scalar mass as functions of $\mu$, on a $12^3 \times 24$ lattice
at $m=0.025$. The curves are the scaling predictions mentioned in the text.}
\label{fig:scalar}
\end{figure}

The third pseudo-Goldstone boson is also a scalar state. It is the linear
combination of the scalar diquark state and the flavour-singlet scalar meson
given in equation~\ref{eqn:scalar}. Here $\alpha$ is unambiguously defined by
the requirement that this state has zero vacuum-expectation value, so we have
calculated the propagator for this state. For $\mu < m_\pi/2$ and $\lambda=0$
this state is a pure diquark state whose propagator is identical to that of
the Goldstone state. Thus we will find both the Goldstone excitation and the
state we want in this propagator. This will also remain true at finite
$\lambda$. Since for finite $\lambda$, there is no phase transition in going
from $\mu < m_\pi/2$ to $\mu > m_\pi/2$, it follows that these 2 states
continue to mix above the transition, although the mixing becomes small for
$\mu >> m_\pi/2$. For this reason we fit our propagator to the form
\begin{equation}
P_S(t) = A \{ \exp[-m_G t] + \exp[-m_G (N_t-t)] \}
       + B \{ \exp[-m_S t] + \exp[-m_S (N_t-t)] \}
\end{equation}
where $m_S > m_G$ is the mass of this scalar state, in addition to the form with
$A=0$, which is appropriate to the $\mu=0$ propagator where $m_G=m_S$ and to
$\mu >> m_\pi/2$ where the lower lying state decouples. For $\mu$ small but
non-zero, we extract this mass by fitting the propagator of the diquark state
obtained by applying the operator $\bar{\chi}\tau_2\bar{\chi}^T$ to the 
vacuum, to the form
\begin{equation}
P_{qq} = A \{ \exp[-m_S t] + \exp[-m_G (N_t-t)] \}.
\end{equation}
This is obtained from the observation that for a pure diquark state (as opposed
to a mixture of a diquark and an antidiquark) at $\lambda=0$, the effect of the
chemical potential is to add $2\mu$ to the effective mass for propagation 
forward in time and subtract $2\mu$ from the effective mass for propagation
backward in time, thus separating the 2 scalar excitations. The masses
obtained from these fits are plotted in figure~\ref{fig:scalar}.

The analysis of section~3, predicts this mass to be the middle mass obtained
from solving the secular equation~\ref{eqn:secular}, while the corresponding
prediction from chiral perturbation theory is the mass of $\tilde{Q}^{\dagger}$
given in equation~\ref{eqn:dispRel} of the appendix. We show these predictions
on the `data' of figure~\ref{fig:scalar}. The agreement is quite good, in
contrast to the other 2 masses. In the low $\mu$ regime, the mass increases as
$m_\pi + 2\mu$, as expected from general arguments. Just above $m_\pi$ the
mass shows a dip past which it resumes its increase, eventually becoming
linear again with the same slope but zero intercept. Remembering that for
$m=0.1$, these masses should be roughly twice those presented here for
$m=0.025$, it is easy to see why they were too difficult to measure with any
precision at that higher quark mass.

Finally we should mention the radial excitation. Even at $\lambda=0$, its mass
is $ > 2.76$ for all $\mu$. Since this is close to the momentum cutoff ($\pi$)
on the lattice, it is not even clear if this should be considered as a real
state. In any case, this mass is too high to be of more than passing interest.
 
\section{Conclusions}

We have simulated 2-colour lattice QCD with one staggered fermion field
corresponding to 4 continuum flavours, in the fundamental representation
(doublet) of the colour group ($SU(2)$), at a finite chemical potential $\mu$,
and quark mass $m=0.025$. As in previous simulations, we have observed the
transition from the normal state to one characterized by a diquark condensate
which spontaneously breaks quark-number. We have found that the
equation-of-state, which describes the dependence of the diquark and chiral
condensates on $\mu$ and the explicit symmetry breaking parameter $\lambda$,
is well approximated by the tree-level approximation to a chiral Lagrangian in
the linear sigma model class. The critical scaling implied by this analysis
indicates that the transition is second order with mean-field critical 
exponents, as expected from chiral perturbation theory analyses through 
next-to-leading order. The measured critical value of $\mu$ is consistent with
$m_\pi/2$ as expected. Applying the predictions of these fits to our earlier
simulations at $m=0.1$ we see evidence that our equation-of-state also gives a
reasonable description of the mass dependence of these condensates. However,
it is clear that saturation effects (a lattice artifact) limit the range of
applicability of this equation for the higher mass, and we suspect that
$m=0.1$ might well be close to the limit of applicability of this
approximation to chiral perturbation theory, if not to the applicability of
chiral perturbation theory itself.

Our equation-of-state also predicts the $\mu$, $\lambda$ and $m$ dependence
of the quark-number density. Here, it appears that the scaling window is
somewhat narrower than that for the condensates. While the effect of saturation
might explain why this density grows faster than the predictions for larger
$\mu$, we suspect that we are seeing the limitations of using tree-level
results from an effective Lagrangian with only one more parameter than the
leading order chiral perturbation theory Lagrangian, to model departures from
tree-level chiral perturbation theory. Use of next-to-leading order chiral
perturbation theory to fit our measurements would require going beyond what
has already been done \cite{STV1}, and going beyond next-to-leading order
would be extremely difficult. In addition, going to higher order in chiral
perturbation theory introduces more parameters and thus reduces its predictive
power. At high enough $\mu$ chiral perturbation theory will break
down. The scale at which chiral perturbation theory
breaks down is given by the pion decay constant $F_\pi$. From the
numerous higher order calculations in chiral perturbation theory, this
breakdown scale can be estimated to be of the order of a few times $F_\pi$. In
our case at $\mu=0$, we know the chiral condensate, the quark
mass and the pion mass. From the Gell-Mann--Oakes--Renner relation, 
we find that $F_\pi\sim0.25$. We therefore expect that chiral
perturbation theory will break down when $\mu\sim1$. However,
at $\mu\sim0.5$, higher-order corrections should already amount to
$20-30\%$. We therefore conclude that it is very doubtful that chiral
perturbation theory can describe our `data' for the largest chemical
potentials used in our study. Note that this $F_\pi$ is the $F_\pi$ for chiral
perturbation theory with the lattice symmetries (see appendix). The continuum
$F_\pi$ is half this value.

Our large $\mu$ data suggests that $j_0$ is increasing consistent with the
expected rise cubic in $\mu$. Since the arguments leading to this prediction
suggest a counting of degrees of freedom identical to that for free quarks,
this suggests that we are beyond the range of chiral perturbation theory, since
accounting for all these degrees of freedom in terms of hadrons requires
including hadrons other than the pseudo-Goldstone bosons. Unfortunately it is
difficult to disentangle such behaviour from the effects of saturation. A
$\mu^3$ increase in $j_0$ is just what is required to keep the saturation
threshold at a constant $\mu$ in lattice units as the lattice spacing is
varied, which is what we observe. In addition, the decrease of the condensate
close to saturation, suggests that the quarks are acting like free quarks,
which makes it even harder to distinguish real effects from saturation induced
artifacts.



The main goal of this project was to measure the spectrum of pseudo-Goldstone
bosons for this theory, to enhance our knowledge of the pattern of symmetry
breaking. Spontaneous breaking of the $U(2)$ symmetry of the staggered lattice
implementation of 2-colour QCD with 1 staggered quark field corresponding to
4 continuum flavours at $m=\lambda=\mu=0$, should give rise to 3 Goldstone
bosons. Fixing $m=0.025$, we have studied the variation of these now
pseudo-Goldstone bosons as functions of $\mu$ for 2 choices of $\lambda$.
Comparison with our earlier simulations at $m=0.1$ gave some indication of
the $m$ dependence of these spectra. We have obtained the predictions from
our linear sigma model effective Lagrangian for these masses, using the
parameters obtained from our fits to the diquark condensate.

For $\mu=0$ our measurements confirm that all 3 pseudo-Goldstone bosons
are degenerate with mass consistent with the expected $M_\pi$. $m_\pi$ is
approximately proportional to $\sqrt{m}$ as predicted by PCAC. The observed
small deviations from PCAC suggest that $m=0.1$ is beyond the range of the
leading order prediction. At $\lambda=0$, the lowest mass state is the diquark
state orthogonal to the condensate. The mass of this state is expected to fall
linearly to zero as $\mu$ is increased to $m_\pi/2$. Above this phase
transition it should remain zero, becoming the Goldstone boson of spontaneously
broken quark number. At small non-zero $\lambda$, our effective Lagrangian
analysis predicts its behaviour. What we have observed is that these
predictions are good up to and through the transition. As $\mu$ is increased
much beyond this value, the measured mass lies consistently below these
predictions. Examining the corresponding predictions for the higher quark mass
$m=0.1$, where the deviation is more severe, suggests that at least some of
this deviation is coming from saturation, a lattice artifact seen when $\mu$
is large enough that the Fermi surface approaches the lattice cutoff.

When $\lambda=0$, we expect that the pseudoscalar pseudo-Goldstone mass will
remain constant as $\mu$ is increased up to $m_\pi/2$. Above this it is
expected to decrease rapidly, approaching zero for large $\mu$. Our measured
values at $m=0.025$ and non-zero $\lambda$ show evidence for such behaviour.
However, the rate of decrease in this mass above the transition is
significantly faster than that predicted by the linear sigma model effective
Lagrangian fits. At $m=0.1$, this decrease is even more precipitous. Here
again, there is evidence to suggest that this more rapid decrease is due to
the effects of saturation. The improvement in going to the smaller mass, where
the onset of saturation occurs at larger $x=2\mu/m-\pi$, supports this
interpretation, and suggests that if the mass were decreased, eventually the
predictions would fit the `data'.

We have noted that the term in the effective Lagrangian which controls the
behaviour of the quark-number density, also contributes to these 
pseudo-Goldstone masses, and does so in a way which would be expected to
make the discrepancies worse for the pseudoscalar state. Hence it is reasonable
to assume that the addition of one extra parameter in going from the chiral
perturbation theory Lagrangian to the linear sigma model Lagrangian, is
inadequate to parameterize all departures from tree-level chiral perturbation
theory. The behaviour of $j_0$ at high $\mu$ can be obtained from analyses
other than those of effective Lagrangians. This suggests that one should
abandon the use of effective chiral Lagrangians designed for small $\mu$ and
small $m_\pi$ and adopt a different approach for large $\mu$.

At $\lambda=0$, the final pseudo-Goldstone mass should increase linearly
with $\mu$ up to $m_\pi/2$, above which it should briefly decrease before
continuing its rise. We see evidence for this behaviour in our measurements
for $m=0.025$. The predictions from effective Lagrangians are in good agreement
with the `data' for this state.

We note that some of the other hadrons which could be expected to contribute
at high $\mu$, are those that would have been pseudo-Goldstone bosons were it
not for the flavour symmetry breaking of the staggered lattice. On the lattice,
the symmetry breaking is $U(2) \rightarrow U(1)$, giving 3 Goldstone bosons.
In the continuum the symmetry breaking is $SU(8) \rightarrow Sp(8)$, which
gives 27 Goldstone bosons. 

To summarize, we have found $m=0.025$ to be small enough to see evidence for
mean-field scaling and to study the spectrum of the 3 pseudo-Goldstone 
excitations. At this quark mass, the pseudo-Goldstone boson masses lie well
below those expected for other `hadrons'. However, saturation, where all
available fermion levels are filled, is still close enough to the phase
transition to make it difficult to disentangle real physics from this lattice
artifact, even though we do find an adequate scaling region for simple
observables. For $m=0.1$ the scaling window is too small to obtain
quantitative information. It would be useful to obtain the full next-to-leading
order analysis of the pseudo-Goldstone spectrum, and the expressions for the
order parameters beyond leading order in $\alpha$. However, this is beyond the
scope of this paper.

\section*{Acknowledgments}

DKS is supported by the US Department of Energy under contract W-31-109-ENG-38.
JBK and DT are supported in part by NSF grant NSF-PHY-0102409. DT is supported
in part by ``Holderbank''-Stiftung. These simulations were performed on the
IBM SPs at NERSC and NPACI.

\appendix
\section{Chiral Perturbation Theory and the Goldstone Spectrum}

In this appendix, we construct chiral perturbation theory for the
symmetry breaking pattern of the staggered fermion action 
(\ref{eqn:lagrangian}) at $m=\lambda=\mu=0$: $U(2)\rightarrow
U(1)$, and we study the spectrum of the three Goldstone excitations. 
These Goldstone modes become massive upon the introduction
of a nonzero quark mass or a nonzero diquark source. They dominate the
physics at low energy. In this appendix we study the spectrum in Chiral 
Perturbation Theory.
This problem is similar to what can be found in the literature. In
\cite{KSTVZ}, chiral perturbation theory for $N_f$ quarks in the
adjoint representation has been constructed. In this case the
symmetry breaking pattern is given by $SU(2N_f)\rightarrow
SO(2N_f)$. Notice that for any number of 
flavors the symmetry breaking pattern of the staggered fermion action
is $U(2N_f)\rightarrow SO(2N_f)$.


Following \cite{KSTVZ,ChPTnf=3}, we construct the low-energy effective
Lagrangian for the Goldstone modes induced by the spontaneous symmetry
breaking $U(2) \rightarrow U(1)$. We find that the effective
Lagrangian is given by
\begin{eqnarray}
  \label{Leff}
  {\cal L}_{\rm eff}=\frac{F^2}2 {\rm Tr} \nabla_\nu \Sigma \nabla_\nu
  \Sigma^\dagger - F^2 M_\pi^2 {\rm Re} {\rm Tr} \hat{M}_\phi \Sigma,
\end{eqnarray}
where $F$ is the pion decay constant, $M_\pi^2=\sqrt{m^2+\lambda^2} \langle
\bar{\psi} \psi\rangle_0/2 F^2$, and   
$\langle\bar{\psi} \psi\rangle_0$ is the quark-antiquark condensate at
$m=\lambda=\mu=0$. 

We have used the same conventions as in \cite{KSTVZ}. These notations 
 were already introduced in section~3, we just need to replace
 $\Sigma_l$ by $\Sigma$ in the expressions that appear in section~3, with
\begin{eqnarray}
\label{Sigma}
  \Sigma=U \bar{\Sigma} U^T,
\end{eqnarray}
where
\begin{eqnarray}
  U=\exp \left( \frac{i\Pi}{2\sqrt{2}F} \right)
 \hspace{.5cm} {\rm with} \hspace{.5cm}
  \Pi=\left(\begin{array}{cc} P_S & Q_R+iQ_I \\ Q_R-iQ_I & P_S
  \end{array}\right),
\end{eqnarray}
and
\begin{eqnarray}
  \bar{\Sigma}=\left(\begin{array}{cc} i \sin \alpha& \cos \alpha \\
      \cos \alpha & i \sin \alpha \end{array}\right),
\end{eqnarray}
corresponds to the minimum of the free energy  with $\alpha$ given by
\begin{eqnarray}
  \label{SPeq}
  4 \mu^2 \cos \alpha \sin \alpha=M_\pi^2 \sin(\alpha-\phi).
\end{eqnarray}

We now turn to the study of the spectrum. 
All the computations made in \cite{KSTVZ} can be easily implemented in
our case, since our effective Lagrangian is very similar to that
studied in \cite{KSTVZ}.

First at $\mu=0$, we find that the mass of the three
pseudo-Goldstone modes is given by $M_\pi$.
At $\mu\neq0$, we can expand the effective Lagrangian (\ref{Leff}) 
to second order
in the Goldstone fields. 

Following \cite{KSTVZ}, we find that the term quadratic in the
Goldstone fields in the effective Lagrangian is given
by
\begin{eqnarray} \label{Leffj}
{\cal L}_{\rm eff}&=&
 {\rm Tr}\left[\left(\partial_\nu Q_R^\dagger\partial_\nu Q_R
+\partial_\nu Q_I^\dagger\partial_\nu Q_I\right)
-4\mu\cos\alpha \left(Q_I^\dagger\partial_0Q_R + Q_R^\dagger\partial_0 Q_I
\right)\right]  \nonumber \\
&&+
M_{\pi}^2 {\rm Tr} \left[
Q_IQ_I^\dagger{\sin\phi\over\sin\alpha}
+ 
Q_RQ_R^\dagger\left(\frac{4\mu^2}{M_\pi^2} \sin^2\alpha +
  {\sin\phi\over\sin\alpha}\right) 
\right] \nonumber\\ 
&&+
{\rm Tr}\left[\partial_\nu P_S \partial_\nu P_S+
P_S^2 M_\pi^2 \left(\frac{4\mu^2}{M_\pi^2} \cos^2\alpha +
  {\sin\phi\over\sin\alpha}\right) \right]. 
\end{eqnarray}

The $Q$ and $Q^\dagger$ modes are mixed. If we call $\tilde{Q}$ and
$\tilde{Q}^\dagger$ the eigenmodes, we find that
the dispersion laws for the different Goldstone modes are given by
\begin{eqnarray}\label{eqn:dispRel}
P_S  &:&  E^2 =  \mbox{\boldmath$p$}^2 +
M_\pi^2 \left(\frac{4\mu^2}{M_\pi^2}\cos^2\alpha 
  +\frac{\sin\phi}{\sin\alpha} \right) ;
\nonumber\\
\tilde Q^\dagger &:& E^2=\mbox{\boldmath$p$}^2 + M_\pi^2
  \frac{\sin\phi}{\sin\alpha} + 2 \mu^2 (1+3 \cos^2\alpha)  \\
 && \hskip 4em + 2 \mu \sqrt{\mu^2 (1+3 \cos^2\alpha)^2+4 \cos^2\alpha \; (
  \mbox{\boldmath$p$}^2+ M_\pi^2 \frac{\sin\phi}{\sin\alpha} )} ;
\nonumber\\
\tilde Q  &:& E^2=\mbox{\boldmath$p$}^2 + M_\pi^2
  \frac{\sin\phi}{\sin\alpha} + 2 \mu^2 (1+3 \cos^2\alpha) \nonumber \\
 && \hskip 4em - 2 \mu \sqrt{\mu^2 (1+3 \cos^2\alpha)^2+4 \cos^2\alpha \; (
  \mbox{\boldmath$p$}^2+ M_\pi^2 \frac{\sin\phi}{\sin\alpha} )}. \nonumber
\end{eqnarray}



\begin{thebibliography}{999}

\bibitem{Shuryak}
R. Rapp, T. Schafer, E.V. Shuryak and M. Velkovsky, Phys. Rev. Lett. {\bf 81},
53 (1998).
\bibitem {Wilczek}
M. Alford, K. Rajagopal and F. Wilczek, Phys. Lett. {\bf B422}, 247 (1998).
\bibitem{Love}
D. Bailin and A. Love, Phys. Rept. {\bf 107}, 325 (1984).
\bibitem{Barrois}
B.~C.~Barrois, Nucl. Phys. B219, 390 (1977)
\bibitem{mu-T}
Z.~Fodor and S.~D.~Katz, Phys. Lett. B534, 87 (2002);
Z.~Fodor and S.~D.~Katz, JHEP 03, 014 (2002);
S.~Choe, {it et al.}, Phys. Rev. D65, 054501 (2002);
C.~R.~Allton, {it et al.}, Phys. Rev. D66, 074507 (2002);
P.~de Forcrand and O.~Philipsen, eprint hep-lat/0205016 (2002);
P.~R.~Crompton, Nucl. Phys. B619, 499 (2001);
P.~R.~Crompton, Nucl. Phys. B626, 228 (2002).
\bibitem{Toublan}
J.B. Kogut, M.A. Stephanov and D. Toublan, Phys. Lett. {\bf B464}, 183 (1999).
\bibitem{KSTVZ}
J.~B.~Kogut, M.~A.~Stephanov, D.~Toublan, J.~J.~Verbaarschot and A.~Zhitnitsky,
Nucl.\ Phys.\ B {\bf 582}, 477 (2000).
\bibitem{STV1}
K.~Splittorff, D.~Toublan and J.~J.~Verbaarschot,
Nucl.\ Phys.\ B {\bf 620}, 290 (2002);
\bibitem{STV2}
Nucl.\ Phys.\ B {\bf 639}, 524 (2002).
\bibitem{SSS}
K.~Splittorff, D.~T.~Son and M.~A.~Stephanov,
Phys.\ Rev.\ D {\bf 64} (2001) 016003.
\bibitem{Vanderheyden}
B.~Vanderheyden and A.~D.~Jackson,
Phys.\ Rev.\ D {\bf 64}, 074016 (2001).
\bibitem{HKLM}
S.~Hands, J.~B.~Kogut, M.~P.~Lombardo and S.~E.~Morrison,
Nucl.\ Phys.\ B {\bf 558}, 327 (1999)
[arXiv:hep-lat/9902034].
\bibitem{HKMS}
S. Hands, J.B. Kogut, S.E. Morrison and D.K. Sinclair, Nucl. Phys. Proc. Suppl.
{\bf94}, 457 (2001).
J.B. Kogut, D.K. Sinclair, S. Hands, and S.E. Morrison, Phys. Rev. D64,
094505 (2001).
\bibitem{KTS1}
J.~B.~Kogut, D.~Toublan and D.~K.~Sinclair,
Phys.\ Lett.\ B {\bf 514}, 77 (2001)
[arXiv:hep-lat/0104010].
\bibitem{KTS2}
J.~B.~Kogut, D.~Toublan and D.~K.~Sinclair,
Nucl.\ Phys.\ B {\bf 642}, 181 (2002)
[arXiv:hep-lat/0205019].
\bibitem{KTS3}
J.~B.~Kogut, D.~Toublan and D.~K.~Sinclair,
arXiv:hep-lat/0208076.
\bibitem{aadgg}
R.~Aloisio, V.~Azcoiti, G.~Di~Carlo, A.~Galante and A.~F.~Grillo, Phys. Lett.
B493, 189 (2000);
R.~Aloisio, V.~Azcoiti, G.~Di~Carlo, A.~Galante and A.~F.~Grillo, Nucl. Phys.
B606, 322 (2001).
\bibitem{t+mu}
A.~Nakamura, Phys. Lett. 149B, 391 (1984);
S.~Muroya, A.~Nakamura and C.~Nonaka, eprint nucl-th/0111082 (2001);
Y.~Liu, {\it et al.}, hep-lat/0009009 (2000);
see also M.~P.~Lombardo, hep-lat/9907025 (1999).
\bibitem{Ecker}
G.~Ecker,
``Chiral perturbation theory,''
CERN-TH-6660-92;
{\it In the proceedings of 4th Hellenic School on Elementary Particle
  Physics, Corfu, Greece, 2-20 Sep 1992 and Lectures given at Cargese
  Summer School on Quantitative Particle Physics, Cargese, 20 Jul-1
  Aug 1992} 
\bibitem{quenched}
J.~B.~Kogut and D.~K.~Sinclair,
Phys.\ Rev.\ D {\bf 66}, 014508 (2002)
[arXiv:hep-lat/0201017].
\bibitem{isospin}
J.~B.~Kogut and D.~K.~Sinclair,
Phys.\ Rev.\ D {\bf 66}, 034505 (2002)
[arXiv:hep-lat/0202028].
\bibitem{sonstep}
D.~T.~Son and M.~A.~Stephanov, Phys. Rev. Lett. 86, 592 (2001);
D.~T.~Son and M.~A.~Stephanov, Yad. Fiz. 64, 899 (2001), Phys. Atom. Nucl. 64,
834 (2001).
\bibitem{ChPTnf=3}
J.~Gasser and H.~Leutwyler,
Nucl.\ Phys.\ B {\bf 250}, 465 (1985).


\end{thebibliography}
\end{document}